\documentclass[12pt, journal, onecolumn, draftclsnofoot]{IEEEtran} 

\usepackage{datetime}
\usepackage{color,soul,cprotect}
\usepackage{mathtools}
\usepackage{booktabs}
\usepackage{multirow} 
\usepackage{cite}
\usepackage{amssymb} 
\usepackage{amsfonts}
\usepackage{amsthm}
\usepackage{verbatim,tikz,pgfplots}

\usetikzlibrary{automata,positioning}
\usetikzlibrary{decorations.pathreplacing,positioning, arrows.meta}

\usepackage{ctable,color,colortbl}
\usepackage{graphicx,epstopdf}
\usepackage{setspace}
\usepackage{algorithm}
\usepackage{algorithmic,eqparbox}

\usepackage{enumerate}
\usepackage[margin=1 in]{geometry}
\usepackage{hyperref}
\usepackage{bbm,bm}
\usepackage{array}
\newcolumntype{?}{!{\vrule width 1pt}}

\usepackage[caption=false,font=normalsize,labelfont=sf,textfont=sf]{subfig}
\usepackage{color}

\makeatletter
\newcommand{\biggg}{\bBigg@{2}}
\newcommand{\Biggg}{\bBigg@{3}}
\newcommand{\Bigggg}{\bBigg@{4}}
\makeatother
\DeclareMathOperator*{\argmax}{argmax}

\newcommand{\figref}[1]{{Fig.}~\ref{#1}}
\newcommand{\tabref}[1]{{Table}~\ref{#1}}
\newcommand{\secref}[1]{{Section}~\ref{#1}}
\newcommand{\algref}[1]{{Algorithm}~\ref{#1}}

\def\bb0{{\mathbb{0}}}

\def\ba{{\mathbf{a}}}
\def\bb{{\mathbf{b}}}

\def\bff{{\mathbf{f}}}

\def\b0{{\mathbf{0}}}

\def\bA{{\mathbf{A}}}

\def\bbE{{\mathbb{E}}}

\def\cA{\mathcal{A}}

\def\cD{\mathcal{D}}

\def\cF{\mathcal{F}}
\def\cG{\mathcal{G}}

\def\cN{\mathcal{N}}

\def\cT{\mathcal{T}}

\def\cW{\mathcal{W}}

\def\sf0{{\mathsf{0}}}

\usepgfplotslibrary{external}

\newcommand{\mytilde}{\raise.17ex\hbox{$\scriptstyle\mathtt{‌​\sim}$}}

\graphicspath{{figure/}{figures/}}

\newcommand{\jj}{\mathrm{j}}

\newcommand{\fRF}{\bm{\mathsf{f}}_{\text{RF}}}

\newcommand{\wRF}{\bm{\mathsf{w}}_{\text{RF}}}

\newcommand{\horizonlen}{M}
\newcommand{\horizonindex}{m}
\newcommand{\subcarriernum}{K}
\newcommand{\subcarrierindex}{k}
\newcommand{\pathnum}{L}
\newcommand{\pathindex}{\ell}
\newcommand{\relayindex}{n}
\newcommand{\modeindex}{n_{\text{mode}}}
\newcommand{\nrelay}{N_{\text{REL}}}

\newcommand{\Nb}{N_{\text{b}}}

\newcommand{\delaytap}{N_{\text{d}}}
\newcommand{\codebooksize}{N_{\text{c}}}

\newcommand{\Nt}{N_{\text{TX}}}
\newcommand{\Nr}{N_{\text{RX}}}

\newcommand{\avgreceivedpower}{G[\horizonindex]}
\newcommand{\noisevector}{\bm{\mathsf{n}}}
\newcommand{\channelmatrix}{\bm{\mathsf{H}}}

\newcommand{\state}{\cT}
\newcommand{\action}{\cA}

\newcommand{\blockagelen}{N_{\text{BL}}}
\newcommand{\noiseSTD}{\sigma_{\text{n}}}

\newcommand{\phinA}{\phi_{\pathindex,\text{A}}}
\newcommand{\phinD}{\phi_{\pathindex,\text{D}}}
\newcommand{\corcoeff}{\sigma_{\text{p}}}
\newcommand{\angularspread}{\sigma_{\text{a}}}
\newcommand{\TXArrayResp}{\ba_{\text{t}}}
\newcommand{\RXArrayResp}{\ba_{\text{r}}}
\newcommand{\thetaA}{\boldsymbol{\theta}_{\text{A,ON}}}
\newcommand{\thetaC}{\boldsymbol{\theta}_{\text{C,ON}}}
\newcommand{\targthetaA}{\boldsymbol{\theta}_{\text{A,TAR}}}
\newcommand{\targthetaC}{\boldsymbol{\theta}_{\text{C,TAR}}}
\newcommand{\taurelay}{\tau_{\text{relay}}}
\newcommand{\taumode}{\tau_{\text{mode}}}
\newcommand{\TXRXchannel}{\channelmatrix_{\text{TX}\rightarrow\text{RX}}}
\newcommand{\TXrelaychannel}{\channelmatrix_{\text{TX}\rightarrow\text{REL}}}
\newcommand{\Txmthrelaychannel}{\channelmatrix_{\text{TX}\rightarrow\text{REL}^{},\relayindex}}
\newcommand{\mthrelayRxchannel}{\channelmatrix_{\text{REL}^{}\rightarrow\text{RX},\relayindex}}
\newcommand{\relayRXchannel}{\channelmatrix_{\text{REL}\rightarrow\text{RX}}}

\newcommand{\optaction}{a_{\text{OPT}}}
\newcommand{\NSS}{N_{\text{SS}}}
\newcommand{\SSperiodicity}{\horizonlen_{\text{SS}}}
\newcommand{\NBT}{N_{\text{BT}}}

\newcommand{\setofrelayindex}{\{0,1,\ldots,N_{\text{REL}}\}}
\newcommand{\BAlen}{\horizonlen_{\text{BA}}}
\newcommand{\DTlen}{\horizonlen_{\text{DT}}}
\newcommand{\IAlendir}{\horizonlen_{\text{IA,direct}}}
\newcommand{\IAlenindir}{\horizonlen_{\text{IA,indirect}}}
\newcommand{\BTlendir}{\horizonlen_{\text{BT,direct}}}

\newcommand{\perfectSEdirectTXRX}{S_{\text{TX}\rightarrow \text{RX},0,\text{p}}}

\newcommand{\measuredSEdirectTXRX}{S_{\text{TX}\rightarrow \text{RX},0}}
\newcommand{\SEdirectTXREL}{S_{\text{TX}\rightarrow \text{REL}^{},\relayindex}}
\newcommand{\SEdirectRELRX}{S_{\text{REL}^{}\rightarrow \text{RX},\relayindex}}
\newcommand{\SEindirectTXRX}{S_{\text{TX}\rightarrow \text{RX},\relayindex}}
\newcommand{\BAlasted}{\horizonindex_{\text{BA}}}
\newcommand{\DTlasted}{\horizonindex_{\text{DT}}}

\newcommand{\channeldelay}{\horizonindex_{\text{d}}}
\newcommand{\batchsize}{B}

\newcommand{\blockagecoeff}{c_{\text{BL},\pathindex}[\horizonindex]}

\title{Joint Relay Selection and Beam Management Based on Deep Reinforcement Learning for Millimeter Wave Vehicular Communication}
\author{
Dohyun Kim, 
Miguel R. Castellanos,~\IEEEmembership{Member, IEEE}, \\ and Robert W. Heath Jr.,~\IEEEmembership{Fellow, IEEE}
\thanks{Dohyun Kim is with the Wireless Networking and Communications Group, the University of Texas at Austin, TX 78712-1687, USA (e-mail: dohyun.kim@utexas.edu). Miguel R. Castellanos and Robert W. Heath Jr. are with the Department of Electrical and Computer Engineering, North Carolina State University, 890 Oval Dr., Raleigh, NC 27606 USA (email: \{mrcastel, rwheathjr\}@ncsu.edu). This work was partially supported by the U.S. Army Research Labs under grant W911NF-19-1-0221 and by the National Science Foundation under grant No. NSF-EECS-2153698.}
}
\begin{document}
\maketitle
\begin{abstract}
Cooperative relays improve reliability and coverage in wireless networks by providing multiple paths for data transmission. Relaying will play an essential role in vehicular networks at higher frequency bands, where mobility and frequent signal blockages cause link outages. To ensure connectivity in a relay-aided vehicular network, the relay selection policy should be designed to efficiently find unblocked relays. Inspired by recent advances in beam management in mobile millimeter wave (mmWave) networks, this paper address the question: \textit{how can the best relay be selected with minimal overhead from beam management?} In this regard, we formulate a sequential decision problem to jointly optimize relay selection and beam management. We propose a joint relay selection and beam management policy based on deep reinforcement learning (DRL) using the Markov property of beam indices and beam measurements. The proposed DRL-based algorithm learns time-varying thresholds that adapt to the dynamic channel conditions and traffic patterns. Numerical experiments demonstrate that the proposed algorithm outperforms baselines without prior channel knowledge. Moreover, the DRL-based algorithm can maintain high spectral efficiency under fast-varying channels.
\end{abstract}

\section{Introduction}

MmWave multiple-input multiple-output (MIMO) communication is a key technology for sensor data sharing to support automation in transportation systems \cite{VaShiBan:MmWave-V2X-Survey:16}. Data sharing between self-driving vehicles can increase the safety of autonomous driving by enabling exchanges of traffic conditions and collision warnings. Safety-critical automated driving applications may require a maximum communication delay of tens-of-milliseconds to prevent catastrophic accidents \cite{GarMolBob:5G_NR_V2X_Tutorial:IEEE21}. Communication at gigabit-per-second data rates will be pivotal to transmit high-resolution data, either raw or processed, from sources such as cameras and radars \cite{KimLiuAng:Cooperative-perception-autonomous-vehicles:ITSM15,ChoVaHea:MmWave-vehicular-massive-automotive-sensing:16}. MmWave MIMO systems can meet the data rate requirements of vehicular networks with beamforming by taking advantage of wide bandwidth communication between 30 and 300 GHz.

Unfortunately, high mobility and frequent blockages in mmWave vehicular networks create a lack of link resilience that may disrupt automotive applications  \cite{TasEgaPia:Modeling-mmWave-highway-vehicular-commuication:VT17}. High mobility systems are subject to fast fading channels, Doppler effects, and frequent handovers. Blockages due to mobile obstacles such as people and cars can induce shadowing losses up to 30-40 dB \cite{GapSamGer:Temporal-effect-mobile-block-urban-mmwave-cellular:VT17}, while blockages due to static objects such as large buildings may result in penetration losses of 40-80 dB \cite{RanRapErk:MmWave-cellular-Challenges:14}. Issues stemming from mobility and blockage can deteriorate the system throughput, and these challenges must be addressed to enable the success of mmWave MIMO networks \cite{MarAguGom:Tech-Model-challenge-mmWave-vehicular-comm:COMMM18}.

Link vulnerability due to mobility can be partially overcome with careful beam management. Though Doppler frequencies are high at mmWave, directional beamforming reduces the effect of Doppler spread by restricting the range of Doppler frequency shifts according to the received beam directions \cite{VaChoHea:Beamwidth_impact_temporal_channel_variation_vehicular_channel:VT16}. While narrow beamwidths can mitigate Doppler spread, narrow codebooks increase the training overhead of exhaustive and hierarchical beam alignment methods. Although prior research has proposed fast beam adaptation in vehicular networks, which addresses the beam alignment overhead \cite{SimKloHol:Online_Context_Aware_ML_mmWAVE_VANET:TNET18,WuCheZha:Fast-mmWave-beam-alignment-correlated-bandit:TCOMM18,PerDelBen:MmWave-V2V-distributed-association-beam-alignment:JSAC17}, most of this work has only considered cellular networks and one-hop transmission links between base stations and vehicles. Few studies have addressed beam alignment overhead in the context of vehicular networks with multi-hop links, despite the benefits of connected vehicles on cooperative decision making such as lane changing and deceleration/acceleration \cite{MaWanYan:AI-application-survey-autonomous-vehicles:20}.

Multi-hop communication, enabled by relaying, can enhance link connectivity by providing multiple transmission paths that can be leveraged to avoid link blockages. In this context, recent studies have shown that a proper selection of unblocked relays can maintain stable data rates with low latency and drop rates \cite{FanTiaZhu:Traffic-aware-vehicle-selection:COMM'L18,ZhaChoLin:mmWave_vehicular_DRL_Relay_Selection_Power_allocation:TCOM'L19,LiXiaGe:Latency-reliability-mmWave-V2V-relay-selection:VT20,GenLiuWan:DDPG_Relay_Selection_Power_allocation:TCOMM'L21}. Recent work on relay selection, however, either has approximated the beamforming gain using an ideal directional antenna pattern \cite{FanTiaZhu:Traffic-aware-vehicle-selection:COMM'L18,ZhaChoLin:mmWave_vehicular_DRL_Relay_Selection_Power_allocation:TCOM'L19,LiXiaGe:Latency-reliability-mmWave-V2V-relay-selection:VT20} or assumed the overhead from beam alignment is negligible \cite{GenLiuWan:DDPG_Relay_Selection_Power_allocation:TCOMM'L21}. Because of this, prior research on relay selection has not accounted for the overhead or the beamforming gain after beam alignment when switching relays.

While a variety of solutions have addressed beam management and relay selection in mmWave MIMO vehicular networks separately \cite{ZhaChoLin:mmWave_vehicular_DRL_Relay_Selection_Power_allocation:TCOM'L19,GenLiuWan:DDPG_Relay_Selection_Power_allocation:TCOMM'L21,SimKloHol:Online_Context_Aware_ML_mmWAVE_VANET:TNET18,BooSurMic:MAB-beam-alignment-tracking-mobile-mmWave:COMM'L19}, the extension to the joint formulation of beam management and relay selection is nontrivial. Beam alignment is needed to establish a robust link when switching to a new relay. The training overhead required for beam alignment, however, may outweigh the benefit of the new relay over the present link. In this context, we develop a DRL-based algorithm that chooses between when to select new relays and when to perform beam management. 

DRL is an online learning method that has been successfully applied to many communication applications, such as network access, caching, and connectivity preservation  \cite{LuoHoaGon:DRL_application_Communications_Survey:COMM19}. In mmWave vehicular networks, DRL has been used for resource allocation and radio access to enhance throughput while maintaining data security \cite{TanKawKat:6G-intelligent-secure-VANET-ML:19}. DRL resolves the exploration-exploitation tradeoff, which appears in many control layer tasks such as dynamic beam selection \cite{SimKloHol:Online_Context_Aware_ML_mmWAVE_VANET:TNET18}, power allocation \cite{ZhaChoLin:mmWave_vehicular_DRL_Relay_Selection_Power_allocation:TCOM'L19}, and handover \cite{VanNguHoa:Handover_RL:TCOM21}. DRL enjoys small control overhead by adaptively balancing between testing new control actions versus choosing the actions deemed to have the maximum expected return according to prior actions deployed. The benefits of DRL make it a suitable approach for solving the joint beam management and relay selection problem. 

In this paper, we propose a DRL algorithm for joint relay selection and beam management that uses beam measurements, which are the rate estimates fed back from the receiver to the transmitter, to decide when to switch relays and when to perform beam alignment. We presume the available relays, which can change over time due to the varying network topology, are identified and at most a two-hop link is allowed. We also assume the communication nodes employ Orthogonal Frequency Division Multiplexing (OFDM), an analog MIMO architecture, codebook-based beamforming, and that the beam measurements are fed back to the transmitter without quantization or overhead. The feedback may be available through a dedicated channel in the sub-6 GHz frequency range or may be sent on the reverse link with reduced coding and spreading. The choice of relay selection or beam management is made by comparing the rate feedback from beam measurements to two adaptive thresholds determined by the algorithm. We use one threshold to determine to either keep or switch the current link, where the links include the direct link and the indirect link through relays. We use the other threshold to decide between data transmission and beam management, which includes initial access, beam tracking, and data transmission \cite{GioPolZor:Beam_Management_3GPP_NR_mmWave_Tutorial:COMM18}. The DRL-based policy uses the best known relay until the performance degrades under the learned threshold, in which case the policy tries out other relays according to beam management procedure. We summarize our contributions as follows:
\begin{enumerate}
\item We formulate a joint relay selection and beam management problem for mmWave MIMO vehicular networks that accounts for the effect of the beam management overhead on the cumulative spectral efficiency. We devise a sequential decision-making model of the joint relay selection and beam management problem, reducing the state space by employing codebook-based beamforming.
\item We propose a DRL-based algorithm to solve the joint relay selection and beam management problem. The proposed algorithm uses the spectral efficiency feedback from the receiver to learn two thresholds, where one threshold corresponds to relay selection and the other to beam management. 
\item We demonstrate the numerical performance between the proposed algorithm versus a baseline with prior knowledge on the channel. The heuristic selects fixed thresholds based on an offline simulation instead of using the DRL algorithm. Note that the heuristic is analgous to the threshold-based relay selection previously studied for cellular device-to-device networks~\cite{WuAtaMas:Threshold_on_SNR_Relaying:TCOM17}. The proposed algorithm is able to outperform the heuristic approach even without the prior knowledge of the channel. Further, we analyze the impact of various system, channel, and beam management parameters on the performance. We find that the proposed DRL-based policy is especially beneficial over baselines under dense vehicular networks with highly-variant channels.  
\end{enumerate}

Relevant studies on relay selection include \cite{PerDelBen:MmWave-V2V-distributed-association-beam-alignment:JSAC17,FanTiaZhu:Traffic-aware-vehicle-selection:COMM'L18,LiXiaGe:Latency-reliability-mmWave-V2V-relay-selection:VT20,HuSchGur:Low-latency-high-reliability-single-relay-selection:VT19,TiaGonChe:Buffer-aided-relay-selection-reduced-packet-delay:VT16,MaChaChe:Relay-assisted-D2D-LTE-A-system:VT17}, which focus on the effective system throughput affected by time overhead. For example, the work in \cite{PerDelBen:MmWave-V2V-distributed-association-beam-alignment:JSAC17} addressed packet overhead and proposed to minimize the average delay of successfully delivered packets. The work in \cite{FanTiaZhu:Traffic-aware-vehicle-selection:COMM'L18} characterized  latency in mmWave vehicular networks as the sum of transmission delay and alignment delay. The work in \cite{LiXiaGe:Latency-reliability-mmWave-V2V-relay-selection:VT20} followed the latency characterization in \cite{FanTiaZhu:Traffic-aware-vehicle-selection:COMM'L18} to maximize the effective rate assuming zero rate is achievable during beam alignment. The beam alignment delay throughout~\cite{PerDelBen:MmWave-V2V-distributed-association-beam-alignment:JSAC17,FanTiaZhu:Traffic-aware-vehicle-selection:COMM'L18,LiXiaGe:Latency-reliability-mmWave-V2V-relay-selection:VT20}, though, is dependent only on the beamwidth. Our work uses the number of training beams and a practical 5G new radio (NR) beam alignment procedure \cite{GioPolZor:Beam_Management_3GPP_NR_mmWave_Tutorial:COMM18} to calculate the overhead induced by both initial access and tracking. In \cite{HuSchGur:Low-latency-high-reliability-single-relay-selection:VT19},an overhead constraint is formulated as a bound on the total broadcasting and relaying time. The overhead has been measured in prior studies on buffer-aided relay selection using the queue length \cite{TiaGonChe:Buffer-aided-relay-selection-reduced-packet-delay:VT16} and packet retransmissions \cite{MaChaChe:Relay-assisted-D2D-LTE-A-system:VT17}. The overhead in \cite{HuSchGur:Low-latency-high-reliability-single-relay-selection:VT19,TiaGonChe:Buffer-aided-relay-selection-reduced-packet-delay:VT16,MaChaChe:Relay-assisted-D2D-LTE-A-system:VT17} does not incorporate the beamforming overhead. Our work penalizes latency due to excessive beam training by assuming exhaustive beam sweeping.

DRL has previously been applied for relay selection in wireless communication networks \cite{SuLuZha:Cooperative_Comm_DRL_Relay_Selection:SENS19,ZhaChoLin:mmWave_vehicular_DRL_Relay_Selection_Power_allocation:TCOM'L19,GenLiuWan:DDPG_Relay_Selection_Power_allocation:TCOMM'L21}. In vehicular networks, DRL has also been applied for simultaneous power level allocation and relay selection. In the line of this work, deep Q-learning (DQL) was used in \cite{ZhaChoLin:mmWave_vehicular_DRL_Relay_Selection_Power_allocation:TCOM'L19} for discrete power allocation to minimize the transmission latency. A deep deterministic policy gradient (DDPG) algorithm for continuous power level allocation to maximize the communication success rate was investigated in  \cite{GenLiuWan:DDPG_Relay_Selection_Power_allocation:TCOMM'L21}. Our paper addresses beam management overhead, where transmit power is fully devoted to a selected relay according to the beam measurement feedback. In this context, \cite{ZhaChoLin:mmWave_vehicular_DRL_Relay_Selection_Power_allocation:TCOM'L19} and \cite{GenLiuWan:DDPG_Relay_Selection_Power_allocation:TCOMM'L21} are complementary to our work. In \cite{SuLuZha:Cooperative_Comm_DRL_Relay_Selection:SENS19}, DRL is applied for relay selection in wireless sensor networks with static nodes using a utility function defined by the system throughput and power usage. Our paper includes mobile nodes in a dynamic mmWave wideband channel and also accounts for the beam training overhead. Our paper also applies DRL with beam measurements as the states instead of the channel matrices, which can greatly improve the runtime because of the smaller state space that facilitates learning. Other online learning algorithms that have been applied to the relay selection problem include the multi-armed bandit framework \cite{ZhaTanWan:MAB-load-balancing-mobile-hierarchical-WSN:20,ZhaLiHan:Adversarial-bandit-relay-selection-acoustic-cooperative-network:21}. Notably, fast beam alignment algorithms based on bandits can exploit environmental awareness \cite{SimKloHol:Online_Context_Aware_ML_mmWAVE_VANET:TNET18}, sparsity of mmWave channels \cite{BooSurMic:MAB-beam-alignment-tracking-mobile-mmWave:COMM'L19}, and correlation structure among beams \cite{WuCheZha:Fast-mmWave-beam-alignment-correlated-bandit:TCOMM18}. Our work assumes exhaustive beam sweeping as in \cite{GioPolZor:Beam_Management_3GPP_NR_mmWave_Tutorial:COMM18}, and we leave the extension to more sophisticated beam alignment algorithms for future work.

The rest of the paper is structured as follows. In \secref{sec:system_model}, we present the system model used to represent the mmWave MIMO vehicular network. In \secref{sec:problem_definition}, we formulate the joint relay selection and beam management problem. In \secref{sec:main_algorithm}, we develop a DRL-based algorithm to solve the joint relay selection and mode selection problem. In \secref{sec:experiments}, we numerically evaluate the proposed algorithm compared to baselines with prior knowledge of the channel. Finally, we conclude the paper in \secref{sec:conclusion}.

We use the following notation throughout this paper: $\bA$ is a matrix, $\ba$ is a vector, $a$ is a scalar, and $\cA$ is a set. We denote $\ba^{\mathrm{T}}$ the transpose of $\ba$, $\ba^{*}$ the conjugate transpose, and $\|\ba\|$ the 2-norm. We denote $\lceil a\rceil$ the ceiling function. We denote $\nabla_{x}$ the gradient with respect to a variable $x$. A scalar random variable $a\sim\cD$ follows distribution $\cD$. We denote the Gaussian distribution $\cN(a,b)$ and the complex Gaussian distribution $\cN_{C}(a,b)$ with mean $a$ and variance $b$.

\section{System model}\label{sec:system_model}

In this section, we describe the system model representing a mmWave vehicular network with V2V communication. We first provide a generic view of the network and beam management procedure in \secref{sec:network_model}. We then describe the signal model in \secref{sec:signal_model}. We outline the beam management procedure in \secref{sec:beam_management}.

\subsection{Network model}\label{sec:network_model}

\begin{figure*}[h!]
	\centering
	\includegraphics[width=5in,draft=false]{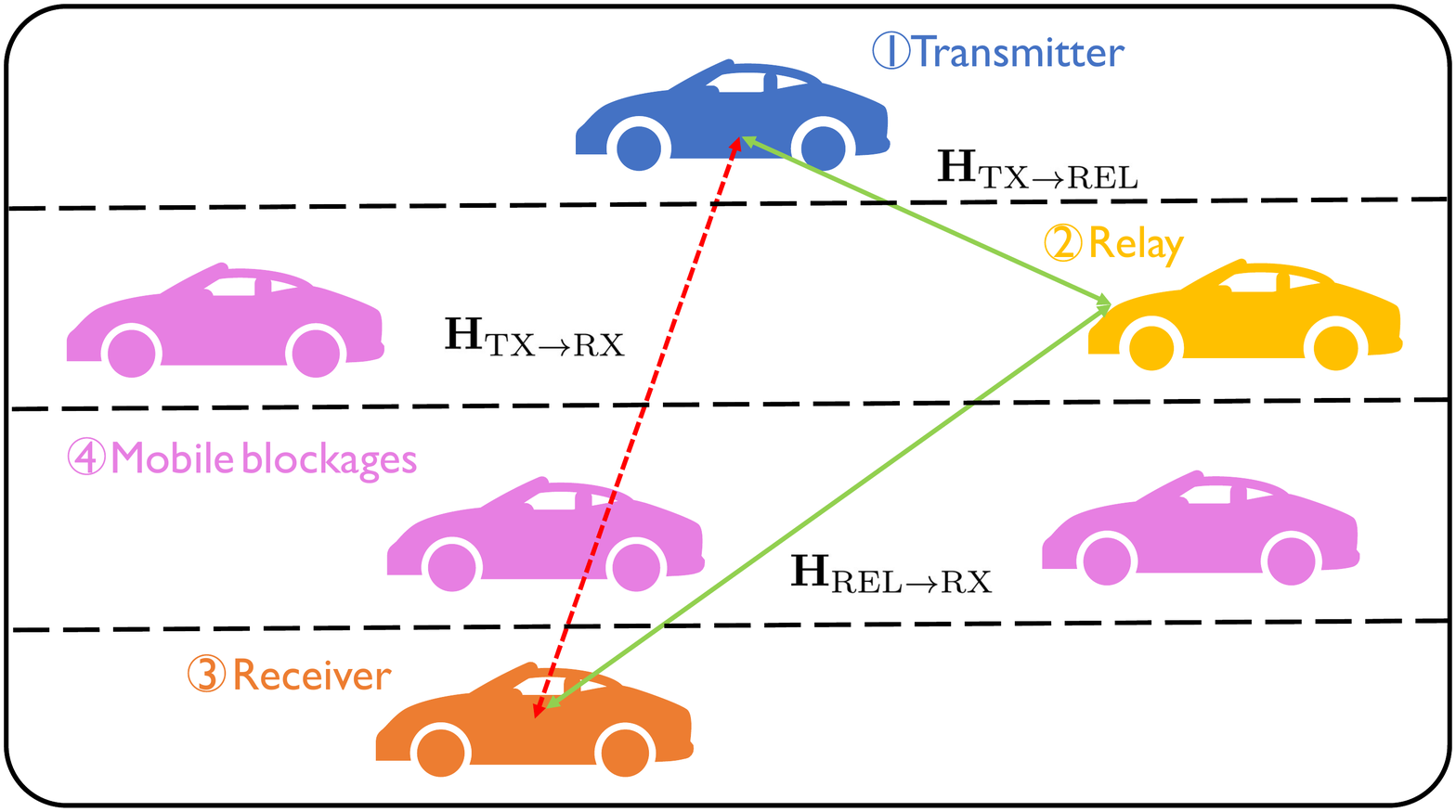}
	\caption{Snapshot illustration of an example system model consisting of four types of vehicles; i) the blue vehicle is the transmitter, ii) the yellow vehicle is an available relay, iii) the orange vehicle is the receiver, and iv) the purple vehicles are mobile blockages. Two-sided arrows indicate vehicular links; solid green links are unblocked and dashed red links are blocked.}
	\label{fig:Highway}
\end{figure*}

Consider a mmWave vehicular network as shown in \figref{fig:Highway}. We assume that the vehicles communicate based on OFDM. The transmitter generates data traffic requested by the receiver, where other vehicles serve as potential relays. The transmitter selects one of two modes, beam alignment or data transmission, for each OFDM frame over the subcarriers and time. We assume the transmitter sends pilots during beam alignment and symbols during data transmission. Whenever the mode is beam alignment, the transmitter performs beam training to send pilots for $\BAlen$ discrete time slots to establish the transmitter-to-receiver link. Otherwise, the transmitter sends data symbols to a single receiver via the transmitter-to-receiver link for $\DTlen$ discrete time slots. This indicates that the sequence of modes can be consecutive beam alignments, consecutive data transmissions, or alternating with an arbitrary number of consecutive modes.

Nearby vehicles can degrade the link quality by blocking the \emph{direct} transmitter-to-receiver path, as shown in \figref{fig:Highway}. We assume the transmitter has already discovered a fixed number $\nrelay$ of nearby relay nodes, given as the set of indices $\setofrelayindex$ where index $0$ denotes the direct transmitter-to-receiver link. Given the indices, the transmitter can establish a two-hop \emph{indirect} transmitter-to-receiver V2V link via the transmitter-to-relay and relay-to-receiver V2V links to overcome the blockage of the direct path. 

\subsection{Signal model}\label{sec:signal_model}

We describe the signal model from the transmitter to the receiver under the data transmission mode. The signal model also applies to other one-hop communication links, such as the transmitter-to-relay and relay-to-receiver link. The signal model under the beam alignment mode is similar to that under the data transmission mode, with the difference that a pilot signal is communicated instead of a data symbol \cite{GioPolZor:Beam_Management_3GPP_NR_mmWave_Tutorial:COMM18}.

We assume an analog beamforming OFDM-MIMO architecture at both the transmitter and receiver. Hybrid and digital architectures allow sweeping over multiple beams simultaneously at the cost of higher energy consumption \cite{GioPolZor:Beam_Management_3GPP_NR_mmWave_Tutorial:COMM18}. Under the analog architecture, the transmitter and receiver communicate via a single data stream. The transmitter consists of $\Nt$ antennas communicating with a receiver with $\Nr$ antennas. We denote $\fRF[\horizonindex]$ the $\Nt\times 1$ complex RF beamformer vector and $\wRF[\horizonindex]$ the $\Nr\times 1$ complex RF combiner vector at time slot $\horizonindex$. We assume frequency flat RF precoder and combiners, such that $\fRF[\horizonindex]$ and $\wRF[\horizonindex]$ are constant over subcarriers, as in \cite{VenGonHea:Optimality-frequency-flat-precoding:COMM'L17}. We assume that the power constraints $\|\fRF[\horizonindex]\|^{2} = 1$ and $\|\wRF[\horizonindex]\|^{2} = 1$, for all $\horizonindex$, on the beamforming vectors $\fRF[\horizonindex]$ and $\wRF[\horizonindex]$. No other hardware-related constraints are assumed. 

We assume a time-varying frequency-selective channel between the transmitter and the receiver. Let us denote $\subcarriernum$ as the number of subcarriers and $\subcarrierindex=1,\ldots,\subcarriernum$ as the subcarrier index. We denote the $\Nr\times\Nt$ channel matrix as $\channelmatrix[\subcarrierindex,\horizonindex]$ between the transmitter and the receiver for each $\subcarrierindex=1,\ldots,\subcarriernum$. The channels used throughout the paper consist of the transmitter-to-receiver channel $\TXRXchannel[\subcarrierindex,\horizonindex]$, transmitter-to-relay channel $\TXrelaychannel[\subcarrierindex,\horizonindex]$, and relay-to-receiver channel $\relayRXchannel[\subcarrierindex,\horizonindex]$, where we omit the subscripts unless needed. We further assume the channel matrix $\channelmatrix[\subcarrierindex,\horizonindex]$ models the small-scale fading, while the averaged received power denoted by $\avgreceivedpower$ represents the large-scale fading \cite{HeaLoz:MIMO-book:18}. Let us also denote the $\Nr\times 1$ independently and identically distributed (IID) $\cN_{C}(0,\noiseSTD^{2})$ noise vector by $\noisevector$. Then, at subcarrier $\subcarrierindex$ and time slot $\horizonindex$, given the complex scalar $\mathsf{s}[\subcarrierindex,\horizonindex]$ of transmitted symbols such that $\bbE[|\mathsf{s}[\subcarrierindex,\horizonindex]|^{2}] = 1$, the processed received signal at subcarrier $\subcarrierindex$ and time slot $\horizonindex$ is \cite{ParAlkHea:Dynamic-subarray-hybrid-precoding-wideband-mmWave:TCOMM17}
\begin{IEEEeqnarray}{lCr}
\mathsf{y}[\subcarrierindex,\horizonindex]
=\sqrt{\avgreceivedpower}\wRF^{*}[\horizonindex]\channelmatrix[\subcarrierindex,\horizonindex]\fRF[\horizonindex]\mathsf{s}[\subcarrierindex,\horizonindex] +
 \wRF^{*}[\horizonindex]\noisevector[\subcarrierindex,\horizonindex].
\label{eq:end_to_end_signal_SU}
\end{IEEEeqnarray}
Note that these normalizations imply that the signal-to-noise-ratio (SNR) prior to beamforming is $\avgreceivedpower/\noiseSTD^{2}$. As the performance metric, we use the instantaneous spectral efficiency~\cite{HeaLoz:MIMO-book:18} averaged over the subcarriers 
\begin{IEEEeqnarray}{lCr}
S(\fRF[\horizonindex],\wRF[\horizonindex],\channelmatrix[\subcarrierindex,\horizonindex])=\frac{1}{\subcarriernum}\sum_{\subcarrierindex=1}^{\subcarriernum}\log_{2}\left(1+\frac{\avgreceivedpower}{\noiseSTD^{2}}|\wRF^{*}[\horizonindex] \channelmatrix[\subcarrierindex,\horizonindex]\fRF[\horizonindex] |^{2}\right).
\label{eq:spectral_efficiency}
\end{IEEEeqnarray}
The receiver can measure the instantaneous spectral efficiency and feed back the beam measurement to the transmitter, as discussed in \secref{sec:beam_management}.
 
\subsection{Beam management procedure}\label{sec:beam_management}

In this section, we outline the codebook-based beam management procedure. We follow a general approach as in commercial mmWave systems like IEEE 802.11ad and 5G. We assume the transmitter and receiver use beams from beam codebooks. We further assume the system employs a feedback mechanism to estimate the spectral efficiency. For simplicity, we assume the feedback is perfect with no quantization and no additional overhead is induced from the feedback procedure. When the receiver successfully decodes one or more successful transmissions, it feeds back the beam measurement to the transmitter. Otherwise, it feeds back a beam measurement of zero to the transmitter. Note that this is is analogous to the automatic repeat-request (ARQ) used in 802.11 standards.

We describe the overall duration of the beam alignment procedure, which is a dominant factor in the beam management overhead. The beam alignment is performed by iterating over predefined beams to aggregate the beam measurements and select the best beam. Each iteration is controlled by synchronization signal (SS) bursts, where a single SS burst consists of multiple SS blocks \cite{GioPolZor:Beam_Management_3GPP_NR_mmWave_Tutorial:COMM18}. Denoting $\NSS$ as the number of SS blocks per burst, the system can examine $\NSS$ pairs of beams when exchanging a single SS burst. Whenever a single SS burst is exchanged, the next SS burst is exchanged after time $\SSperiodicity$ slots, which we denote as the periodicity of SS bursts. When beam alignment starts at time $\horizonindex$, the first beam pair in the SS burst is exchanged at time $\horizonindex+\lceil\SSperiodicity/\NSS\rceil$, the second beam pair at time $\horizonindex+2\lceil\SSperiodicity/\NSS\rceil$, continuing up to the last beam pair at time $\horizonindex+\NSS\lceil\SSperiodicity/\NSS\rceil$. The duration of the beam alignment period depends on the number of beam pairs that should be examined, which can be categorized into four cases depending on the mode and the number of hops. The mode can be either initial access or beam tracking. The direct link has one hop, and the indirect link has two hops. Let us denote the transmitter codebook with size $\codebooksize$ as $\cF = \{\bm{\mathsf{f}}_{1},\bm{\mathsf{f}}_{2},\ldots,\bm{\mathsf{f}}_{\codebooksize}\}$, and similarly the receiver codebook as $\cW$ and $\relayindex$th relay codebook as $\cG_{\relayindex}$. For initial access via direct link, the duration of beam alignment can be expressed as 
\begin{IEEEeqnarray}{lCr}
	\IAlendir = \SSperiodicity\left\lceil \frac{|\cF|\cdot|\cW|}{\NSS} \right\rceil,
	\label{eq:CElen_initial_access_direct}
\end{IEEEeqnarray}
due to the exhaustive beam sweeping over $\cF \times \cW$. Let us denote $\NBT$ as the number of best beams fed back to the transmitter from the receiver during beam tracking. Unlike in initial access where  $|\cF|\cdot|\cW|$ beams are swept, only $\NBT<<|\cF|\cdot|\cW|$ beams are processed in beam tracking. The duration of the beam alignment period for beam tracking via direct link is
\begin{IEEEeqnarray}{lCr}
	\BTlendir = \SSperiodicity\left\lceil \frac{\NBT}{\NSS} \right\rceil.
	\label{eq:CElen_beam_tracking_direct}
\end{IEEEeqnarray}
For simplicity, let us assume perfect time synchronization between the transmitter and the relay. Then, the duration of the beam alignment procedure is 
\begin{IEEEeqnarray}{lCr}
	\IAlenindir = \SSperiodicity\left\lceil \frac{|\cF|\cdot|\cG_{\relayindex}|}{\NSS} \right\rceil+ 
	\SSperiodicity\left\lceil \frac{|\cG_{\relayindex}|\cdot|\cW|}{\NSS} \right\rceil,
	\label{eq:CElen_initial_access_indirect}	
\end{IEEEeqnarray}
for initial access via indirect link and
\begin{IEEEeqnarray}{lCr}
	\BTlendir = 2\SSperiodicity\left\lceil \frac{\NBT}{\NSS} \right\rceil,
	\label{eq:CElen_beam_tracking_indirect}
\end{IEEEeqnarray}
for beam tracking. The indirect link has a longer beam alignment period than the direct link both for initial access and beam tracking. Nonetheless, the effective spectral efficiency accounting the beamforming overhead may be high in the indirect link due to blockage of the direct link. 

During beam alignment, the transmitter and the receiver search for the best transmit and receive beam pair that maximizes SNR \cite{GioPolZor:Beam_Management_3GPP_NR_mmWave_Tutorial:COMM18}. Due to the exhaustive beam sweeping procedure, beam indices are swept sequentially over time. Let us denote the time slot when codebook indices $(i_{\cF},i_{\cW})$ are being swept as
\begin{IEEEeqnarray}{lCr}
\channeldelay(i_{\cF},i_{\cW})=\left\lceil \frac{\codebooksize(i_{\cF}-1)+i_{\cW}}{\NSS}\right\rceil,
\label{eq:beam_alignment_residual_time}
\end{IEEEeqnarray}
where the subscript d shows the delay due to the exhaustive beam sweeping is accounted. When beam alignment ends at time slot $\horizonindex$, the system obtains the beamforming vectors 
\begin{IEEEeqnarray}{lCr}
(\bm{\mathsf{f}}_{\text{d},i_{\cF}}[\horizonindex],\bm{\mathsf{w}}_{\text{d},i_{\cW}}[\horizonindex]) 
= \argmax_{i_{\cF}\in\cF,i_{\cW}\in\cW} S(\bm{\mathsf{f}}_{i_{\cF}}[\horizonindex],\bm{\mathsf{w}}_{i_{\cW}}[\horizonindex],\TXRXchannel[\horizonindex-\BAlen+\channeldelay(i_{\cF},i_{\cW})]),
\label{eq:beam_sweeping_return_index}
\end{IEEEeqnarray}
and the achievable spectral efficiency is given by
\begin{IEEEeqnarray}{lCr}
\perfectSEdirectTXRX[\horizonindex] = \frac{1}{\subcarriernum}\sum_{\subcarrierindex=1}^{\subcarriernum}\log _{2}\left(1+\frac{\avgreceivedpower}{\noiseSTD^{2}} \bigg|\bm{\mathsf{w}}^{*}_{\text{d},i_{\cW}}[\horizonindex] \TXRXchannel[\subcarrierindex,\horizonindex]\bm{\mathsf{f}}_{\text{d},i_{\cF}}[\horizonindex]\bigg|^{2}\right),
\label{eq:beam_sweeping_spectral_efficiency_direct}
\end{IEEEeqnarray}
where the subscript $0$ indicates using the direct link. The subscript $\text{p}$ indicates no measurement error is included in \eqref{eq:beam_sweeping_spectral_efficiency_direct}. 

To incorporate measurement error, we express the beam measurement assuming the system uses MMSE estimator for the effective channel under a rectangular Doppler spectrum as in \cite[Sec. 4.8]{HeaLoz:MIMO-book:18}. As the MMSE estimator can be obtained in terms of the ratio of pilots per symbol transmission, we count the number of pilots over time and frequency frames between data transmission modes. For every block between data transmission modes, in this context, we denote the varying ratio of pilots as $\beta$ and the total number of OFDM frames as $\Nb$. Then, the MMSE can be written as
\begin{IEEEeqnarray}{lCr}
\text{MMSE}=\frac{1}{1+\beta\Nb\text{SNR}},
\label{eq:MMSE_value}
\end{IEEEeqnarray}
and the effective SNR as 
\begin{IEEEeqnarray}{lCr}
\text{SNR}_{\text{eff}}=\frac{\text{SNR}(1-\text{MMSE})}{1+\text{SNR}\cdot\text{MMSE}}.
\label{eq:effective_SNR}
\end{IEEEeqnarray}
The estimated spectral efficiency, fed back from the receiver to the transmitter as a beam measurement, is 
\begin{IEEEeqnarray}{lCr}
\measuredSEdirectTXRX[\horizonindex] = \frac{1}{\subcarriernum}\sum_{\subcarrierindex=1}^{\subcarriernum}\log _{2}\left(1+\text{SNR}_{\text{eff}} \bigg|\bm{\mathsf{w}}^{*}_{\text{d},i_{\cW}}[\horizonindex] \TXRXchannel[\subcarrierindex,\horizonindex]\bm{\mathsf{f}}_{\text{d},i_{\cF}}[\horizonindex]\bigg|^{2}\right),
\label{eq:beam_measurement}
\end{IEEEeqnarray} when the symbol is being sent at time slot $\horizonindex$ and zero during beam management.
We similarly define the estimated spectral efficiency $\SEdirectTXREL$ and $\SEdirectRELRX$ through transmitter-to-relay and relay-to-receiver link. For $\SEdirectTXREL$, the codebook pair $(\cF,\cW)$ is replaced by $(\cF,\cG_{\relayindex})$ and the channel $\TXRXchannel[\horizonindex]$ with $\Txmthrelaychannel[\horizonindex]$. For $\SEdirectRELRX$, the codebook pair $(\cF,\cW)$ is replaced by $(\cG_{\relayindex},\cW)$ and the channel $\TXRXchannel[\horizonindex]$ with $\mthrelayRxchannel[\horizonindex]$. We replace the subscript $0$ with $\relayindex$ for the transmitter-to-relay and the relay-to-receiver link to indicate using the $\relayindex$th link. The overall spectral efficiency of the two-hop indirect path is
\begin{IEEEeqnarray}{lCr}
\SEindirectTXRX[\horizonindex] = \frac{\SEdirectTXREL[\horizonindex]\SEdirectRELRX[\horizonindex]}{\SEdirectTXREL[\horizonindex]+\SEdirectRELRX[\horizonindex]},
\label{eq:beam_sweeping_spectral_efficiency_indirect}
\end{IEEEeqnarray}
following the optimal time resource allocation for decode-and-forward relaying as in \cite{LiuYanYan:SE-analysis-two-hop-relay:VT13}. The beam measurement of the transmitter-to-relay and relay-to-receiver link may be individually available to the transmitter via the relay-to-transmitter and the receiver-to-transmitter feedback channel.   

\section{Formulating the joint relay selection and beam management problem}\label{sec:problem_definition}

In this section, we formulate the joint relay selection and beam management problem for the mmWave MIMO vehicular network from the perspective of sequential decision theory. Based on this formulation, we discuss how to choose actions for each time steps. To do this, we devise a Markov Decision Process (MDP), which is a well-studied model for sequential decision making. 

The transmitter aims to maximize the data rate by selecting the best relay and beam at each time slot. We say that the transmitter needs to decide \emph{actions} $\action[\horizonindex]$ for each time slot. The actions consist of a chosen relay index $\relayindex[\horizonindex]\in\setofrelayindex$ and a beam management mode $\modeindex[\horizonindex]\in\{0,1\}$ which dictates whether to perform beam alignment or data transmission. We set $\modeindex=1$ to indicate data transmission and $\modeindex=0$ to indicate beam alignment.

The optimal set of actions are selected to maximize the running average of the spectral efficiency over $\horizonlen$ time slots. We assume a finite $\horizonlen$ to ensure the sum of spectral efficiency is bounded, as in other sequntial decision formulations in wireless applications \cite{LuoHoaGon:DRL_application_Communications_Survey:COMM19}. On top of the spectral efficiency depending on the channel and beamforming vectors, as given in \eqref{eq:beam_measurement}, the action affects the spectral efficiency due to the beam management procedure. In this context, we use a binary variable $c(\action[\horizonindex])$ to express the effect of the actions on the spectral efficiency. We set $c(\action[\horizonindex])=1$ when the action is data transmission and $c(\action[\horizonindex])=0$ when the action is beam alignment. Then, the optimization problem for maximizing the cumulative spectral efficiency can be written as
\begin{IEEEeqnarray}{lCr}
\max_{\{a[\horizonindex]\}}  \sum_{\horizonindex=1}^{\horizonlen}\sum_{\relayindex=0}^{\nrelay}\bigg(c(\action[\horizonindex])\SEindirectTXRX[\horizonindex]\bigg).
\label{eq:cumulative_rate_optimization}
\end{IEEEeqnarray}

We first analyze a genie-aided policy to approach \eqref{eq:cumulative_rate_optimization}. At time slot $\horizonindex$, suppose the achievable spectral efficiency $\SEindirectTXRX[\horizonindex]$ is known for all $\relayindex$. In this case, the optimal solution $\optaction[\horizonindex]$ of \eqref{eq:cumulative_rate_optimization} is selecting the relay index $\relayindex[\horizonindex]=\argmax_{\relayindex}\SEindirectTXRX[\horizonindex]$ with the mode $\modeindex[\horizonindex]=1$. Note that the value obtained by $\optaction$ is the expected upper bound of the system's performance. 

The system is limited from achieving the performance of the genie-aided policy due to the tradeoff between the performance obtained from frequent beam alignment versus frequent data transmission. On one hand, frequent beam alignment is necessary due to the fast varying channel. On the other hand, frequent data transmission is required to realize the spectral efficiency. The tradeoff can be also explained in terms of the objective in \eqref{eq:cumulative_rate_optimization}. Frequent beam alignment can improve the accuracy of rate feedback leading to a higher $\SEindirectTXRX[\horizonindex]$ at the expense of the coefficient set to $c(\action[\horizonindex])=0$. Conversely, frequent data transmission can achieve the coefficient $c(\action[\horizonindex])=1$ at the cost of a lower $\SEindirectTXRX[\horizonindex]$ due to beam misalignment. 

The system can address the performance tradeoff between beam alignment versus data transmission using sequential decision theory. Following the approach taken in sequential decision making formulations in wireless communication applications \cite{LuoHoaGon:DRL_application_Communications_Survey:COMM19}, we assume an MDP as the learning model for \eqref{eq:cumulative_rate_optimization}. The three components that must be specified in an MDP are the states, actions, and the reward:
\begin{itemize}
\item \textit{States:} The system state of interest is determined by the channel realizations. In codebook-based directional beamforming, the beam indices \eqref{eq:beam_sweeping_return_index} and measurements \eqref{eq:beam_measurement} can substitute the channel information \cite{NohZolLov:Multi-resolution-codebook-mmWave-beam-alignment:TCOMM17}.
Accordingly, we define the link vector of the communication link via the $\relayindex$th relay as
\begin{IEEEeqnarray}{lCr}
\bb_{\relayindex}[\horizonindex] = \left[i_{\cF,\text{OPT}}[\horizonindex], i_{\cG_{\relayindex},\text{OPT}}[\horizonindex], \SEdirectTXREL[\horizonindex]\right].
\label{eq:beam_vector}
\end{IEEEeqnarray}
The state can then be represented as 
\begin{IEEEeqnarray}{lCr}
\state[\horizonindex] =  \{\bb_{0}[\horizonindex],\ldots,\bb_{\nrelay}[\horizonindex]\},
\label{eq:state_vector}
\end{IEEEeqnarray}
which consists of the link vectors for all relay indices. 
\item \textit{Actions}: The action of the transmitter is the decision variable in the optimization problem \eqref{eq:cumulative_rate_optimization}. Though discrete actions can be used, continuous actions are often preferred in wireless applications due to scalability \cite{LuoHoaGon:DRL_application_Communications_Survey:COMM19}. We follow this approach and defer the readers to \secref{sec:algorithm-threshold-policy} for the specification of the continuous action.
\item \textit{Reward:} The reward is designed to maximize the objective in \eqref{eq:cumulative_rate_optimization}, which can be represented as
\begin{IEEEeqnarray}{lCr}
r(\state[\horizonindex],\action[\horizonindex])=\sum_{\relayindex=0}^{\nrelay}\bigg(c(\action[\horizonindex])\SEindirectTXRX[\horizonindex]\bigg).
\label{eq:reward}
\end{IEEEeqnarray}
Note that we follow the typical approach of choosing the reward as the objective at time index $\horizonindex$ \cite{LuoHoaGon:DRL_application_Communications_Survey:COMM19}.
\end{itemize}

\section{Policy design for joint relay selection and beam management} \label{sec:main_algorithm}

In this section, we develop algorithms to solve the joint relay selection and beam management in mmWave MIMO vehicular networks. We develop a DRL-based algorithm based on a pure threshold policy \cite{WuAtaMas:Threshold_on_SNR_Relaying:TCOM17,BitEftKan:V2V-relaying-outdated-CSI-1D-example:VT17}. In \secref{sec:algorithm-threshold-policy}, we first describe a threshold-based heuristic (\algref{alg:threshold-based-heuristic}) with fixed $\tau_{\textrm{relay}}$ and $\tau_{\textrm{mode}}$ that determine the relay index and mode. We then specify the proposed DRL-based policy, as in \algref{alg:DRL-policy}, which applies DRL based on a policy gradient approach to learn the thresholds and solve the joint relay selection and beam management in \secref{sec:learning_algorithm}. 

\subsection{Threshold-based heuristic}\label{sec:algorithm-threshold-policy}

Threshold-based policies with one threshold have been studied for relay selection \cite{WuAtaMas:Threshold_on_SNR_Relaying:TCOM17,BitEftKan:V2V-relaying-outdated-CSI-1D-example:VT17}. One threshold is sufficient for relay selection, as it can represent one of two behaviors: to either keep the relay or switch. For example, the receiver may switch relays if the estimated received SNR of the current link is below that of the best relay and hold otherwise \cite{BitEftKan:V2V-relaying-outdated-CSI-1D-example:VT17}. With more behaviors to model, however, additional thresholds may be required. For example, threshold-based policies for data transmission through a Gilbert-Eilliot channel often required two separate thresholds to determine to whether send data, wait, or measure the channel \cite{LaoTon:Betting-on-GE-channel:TCOMM10}.

We follow the threshold-based policies as in \cite{LaoTon:Betting-on-GE-channel:TCOMM10} to use thresholds as actions. Two continuous thresholds $\taurelay$ and $\taumode$ are defined such that the action can be represented as 
\begin{IEEEeqnarray}{lCr}
\action[\horizonindex] = \{\taurelay,\taumode\}.
\label{eq:action_vector}
\end{IEEEeqnarray}
The transmitter compares the rate feedback in \eqref{eq:beam_measurement} to the thresholds and then chooses one of the following three behaviors: optimistic, opportunistic, and pessimistic action. When the transmitter is optimistic, believing that the channel is in an unblocked state with high achievable spectral efficiency, it keeps both the relay index and mode. When the transmitter is opportunistic, believing that the channel is in an unblocked state but with a low achievable spectral efficiency, it keeps the relay index but sets the mode to beam tracking. When the transmitter is pessimistic, believing the channel is in a blocked state, it changes the relay index and also sets the mode to beam alignment. We assume $\taurelay<\taumode$ due to the rate of blocked channels being worse than that of the unblocked and bad channels. The belief of the transmitter regarding the channel is determined by the beam measurements in \eqref{eq:beam_measurement}. For a given beam measurement $S$ of the current link, the transmitter takes the optimistic action if $S>\taumode$, the opportunistic action if $\taumode>S>\taurelay$, or the pessimistic action if $\taurelay>S$.

The pseudocode of the proposed threshold-based heuristic is given in \algref{alg:threshold-based-heuristic}. The algorithm requires the thresholds $\taurelay$ and $\taumode$ as fixed inputs. The algorithm is similar to a state transition matrix. It takes $\relayindex[\horizonindex]$, mode $\modeindex[\horizonindex]$, and link vectors $\bb_{0}[\horizonindex],\ldots,\bb_{\nrelay}[\horizonindex]$ at the $\horizonindex$th time slot to obtain $\state[\horizonindex+1]$. Due to the duration of beam management, the algorithm may need to continue the mode $\modeindex[\horizonindex]$ over multiple time slots. To do this, the algorithm tracks how long the current beam management mode has lasted using $\BAlasted[\horizonindex]$ and $\DTlasted[\horizonindex]$. The variable $\BAlasted[\horizonindex]$ can be thought as the number of beam indices swept in the current beam alignment mode \eqref{eq:beam_alignment_residual_time}. The variable $\DTlasted[\horizonindex]$ relates to the number of time slots spent in the current data transmission. At the end of each beam management mode, when $\BAlasted=\BAlen$ or $\DTlasted=\DTlen$, the algorithm updates the relay index and beam management mode depending on the transmitter's belief of the channel.
\begin{algorithm}[]
\caption{Threshold-based heuristic for joint relay selection and beam management problem}
\label{alg:threshold-based-heuristic}
\begin{algorithmic}[1]
\STATE{Input: threshold $\taumode$ on mode selection, threshold $\taurelay$ on relay selection, current time slot index $k$, current relay index $\relayindex[\horizonindex]$, current mode $\modeindex[\horizonindex]$, and current link vectors $\bb_{0}[\horizonindex],\ldots,\bb_{\nrelay}[\horizonindex]$}		
\IF[Beam alignment]{$\modeindex[\horizonindex]=0$}
\STATE{$S[\horizonindex]=0$}
\IF{$\BAlasted[\horizonindex]<\BAlen[\horizonindex]$}
\STATE{$\modeindex[\horizonindex+1]=0$}
\STATE{Update $\BAlasted[\horizonindex+1]=\BAlasted[\horizonindex]+1$ }
\ELSE
\STATE{Update beam indices $\bb_{\relayindex[\horizonindex]}[\horizonindex+1]$ according to \eqref{eq:beam_sweeping_return_index} } \label{algline:heursitic-beam-index-update}
\STATE{$\modeindex[\horizonindex]=1$}	\label{algline:heursitic-continue-channel-estimation}
\STATE{$\BAlasted[\horizonindex+1]=1$}
\ENDIF
\ELSE[Data transmission] 
\STATE{Set measured spectral efficiency  $S[\horizonindex]$ according to $\bb_{\relayindex[\horizonindex]}[\horizonindex]$} 
\IF{$\DTlasted[\horizonindex]<\DTlen$}
\STATE{$\modeindex[\horizonindex+1]=1$}
\STATE{Update $\DTlasted[\horizonindex+1]=\DTlasted[\horizonindex]+1$ }
\ELSE
\IF{$S[\horizonindex]<\taurelay$}
\STATE{$\relayindex[\horizonindex+1] =\argmax_{\relayindex\in\{0,1,\ldots,\relayindex[\horizonindex]-1,\relayindex[\horizonindex]+1,\ldots,\nrelay\}}\SEindirectTXRX[\horizonindex]$}
\STATE{$\modeindex[\horizonindex]=0$}
\ELSIF{$S[\horizonindex] <\taumode$}
\STATE{$\modeindex[\horizonindex]=0$} \label{algline:heursitic-tau-mode}
\ENDIF
\STATE{$\DTlasted[\horizonindex+1]=1$}
\ENDIF
\ENDIF
\STATE{Output: relay index $\relayindex[\horizonindex+1]$, mode $\modeindex[\horizonindex+1]$, link vectors $\bb_{0}[\horizonindex+1],\ldots,\bb_{M}[\horizonindex+1]$, and measured spectral efficiency $S[\horizonindex]$}
\end{algorithmic}
\end{algorithm}

To deploy the threshold-based heuristic, the thresholds $\taurelay$ and $\taumode$ are required as inputs. In practice, test results over varying $\taurelay$ and $\taumode$ may be compared to choose the thresholds that provide the highest spectral efficiency. Considering dense vehicular networks with complex and dynamic traffic patterns, the thresholds need to be computed efficiently both in terms of data and time resources \cite{SimKloHol:Online_Context_Aware_ML_mmWAVE_VANET:TNET18}. For this reason, we apply DRL to find the thresholds with short training time and without offline data.

\subsection{Learning algorithm}\label{sec:learning_algorithm}

DRL algorithms aim to find the sequence of actions that maximize the cumulative reward by training neural networks through trial-and-error. At each iteration an action is determined according to the output of the neural networks. The action is deployed on the environment resulting in a reward. The reward is then used to update the weights of neural networks, which will determine the next action.

The following fundamental aspects are involved in the design of the DRL algorithms: the policy $\mu$ and the Q-function $Q$. The policy is a mapping from the state space to the action space, such that $\action=\mu(\state)$. The aim of DRL is typically formulated as finding the best policy. The Q-function $Q(\state,\action)$ is a measure of the expected reward from a state-action pair followed by the state-action pairs induced by the optimal policy. The Q-function $Q(\state,\action)$ is often useful for policy search problems due to two properties: it provides a straightforward way to find the optimal policy $\mu^{\text{OPT}}(s)=\argmax_{a}Q(\state,\action)$, and it can be computed with Bellman updates \cite{SutBar:RL-Book:18}.

We use DDPG~\cite{LilHunPri:DDPG:15}, which is a DRL algorithm that trains both the policy and Q with neural networks, to solve the joint relay selection and beam management problem. It trains an actor $\thetaA$ that takes states as inputs and actions as outputs. The actor network accordingly yields the policy $\mu_{\thetaA}$. DDPG also trains a critic $\thetaC$ that takes state-action pairs as inputs and Q values as outputs. The critic network represents the Q-function $Q(\cdot|\thetaC)$. For stable learning, DDPG reserves the delayed copy of $\thetaA$ and $\thetaC$ as the target networks $\targthetaA$ and $\targthetaC$.

DDPG is a suitable algorithm for the joint relay selection and beam management, as in other wireless applications, due to its fast convergence and capability of handling continuous action spaces \cite{LuoHoaGon:DRL_application_Communications_Survey:COMM19}. We introduce the updating rule for the neural networks in DDPG. Let us denote the replay buffer as $\cD$. Each element in the replay buffer is a tuple consisting of state, action, reward, and successor state. The tuple $(\state[\horizonindex],\action[\horizonindex],r[\horizonindex],\state[\horizonindex+1])$ is denoted as a trajectory, referring to the deployment history. A $B$-element minibatch, which consist of trajectories randomly sampled with replacement from $\cD$, is used for updating the online actor and critic networks. Specifically, $\thetaC$ is updated by minimizing the loss
\begin{IEEEeqnarray}{lCr}
L = \frac{1}{\batchsize}\sum_{\horizonindex'}\bigg((r[\horizonindex'] + \gamma Q(\state[\horizonindex'+1],\mu_{\targthetaA}(\state[\horizonindex'+1])|\targthetaC) \nonumber\\  - Q(\state[\horizonindex'],\action[\horizonindex']|\thetaC) )^{2} \bigg).
\label{eq:DDPG_loss_function}
\end{IEEEeqnarray}  
Meanwhile, the sampled policy gradient, which updates $\thetaA$, is given as 
\begin{IEEEeqnarray}{lCr}
\sum_{\horizonindex'}\frac{1}{\batchsize}\bigg(\nabla_{\action}Q(\state,\action|\thetaC)|_{\state=\state[\horizonindex'],\action=\mu_{\thetaA}(\state[\horizonindex'])}
\times\nabla_{\thetaA}\mu_{\thetaA}(\state)|_{\state=\state[\horizonindex']}\bigg). \label{eq:DDPG_actor_network_sampled_policy_gradient}
\end{IEEEeqnarray}
The target networks are slowly updated from the online networks, where the parameter $\eta<<1$ controls the variance of the target networks:
\begin{IEEEeqnarray}{lCr}
\targthetaA &\leftarrow \eta\thetaA+(1-\eta)\targthetaA, \nonumber\\
\targthetaC &\leftarrow \eta\thetaC+(1-\eta)\targthetaC.
\label{eq:DDPG_target_from_online}
\end{IEEEeqnarray}
The parameter $\eta$ can be used to suppress the overestimation of the Q-values \cite{FujHooMeg:TD3:ICML18}.

\begin{figure}
\centering
\includegraphics[width=0.5\columnwidth,draft=false]{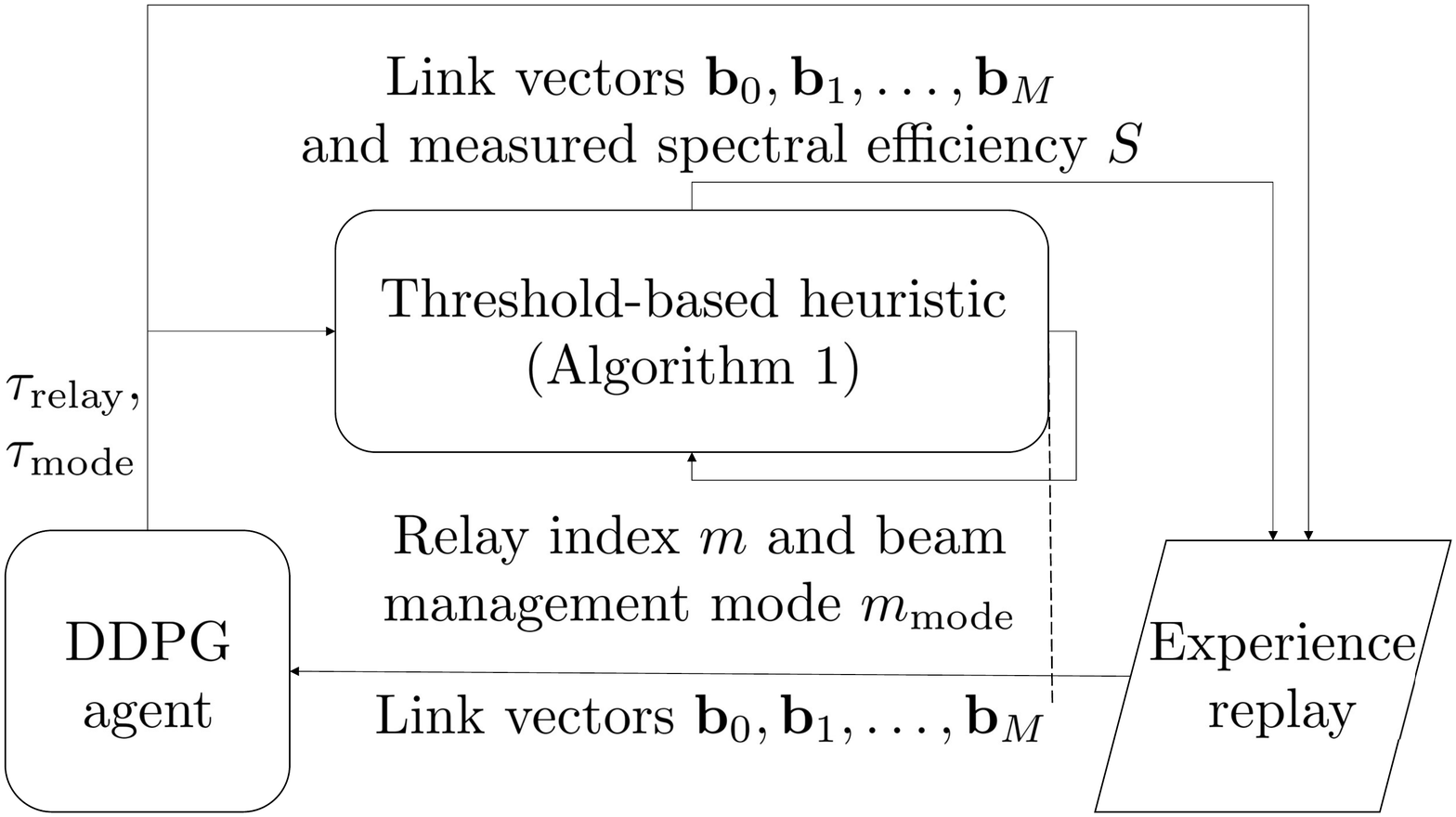}
\caption{Flowchart of the proposed DRL-based joint relay selection and beam management algorithm. The threshold-based heuristic (\algref{alg:threshold-based-heuristic}) serves as the environment in each iteration.}
\label{fig:Algorithm-flowchart}
\end{figure}

Implementing DDPG for joint relay selection and beam management, the following steps are repeated for the time slots $\horizonindex=1,\ldots,\horizonlen$:
\begin{enumerate}
\item Select the thresholds $\taurelay[\horizonindex]$ and $\taumode[\horizonindex]$ according to the online actor network $\targthetaA$ and exploration noise distribution $\cN$, where the default exploration noise is the Ornstein-Uhlenbeck noise. 
\item Deploy \algref{alg:threshold-based-heuristic} with the inputs $\taurelay[\horizonindex]$, $\taumode[\horizonindex]$, $\bb_{0}[\horizonindex],\ldots,\bb_{\nrelay}[\horizonindex]$, $I[\horizonindex]$, $\relayindex[\horizonindex]$, and $\modeindex[\horizonindex]$. As a result, obtain the successive $\bb_{0}[\horizonindex+1],\ldots,\bb_{\nrelay}[\horizonindex+1]$, $\relayindex[\horizonindex+1]$, $\modeindex[\horizonindex+1]$, and $S[\horizonindex]$.
\item Append the current state action pair to the successor state and reward pair to accumulate transition $(\state[\horizonindex],\action[\horizonindex],r[\horizonindex],\state[\horizonindex+1])$ in replay buffer $\cD$.
\item Update the online actor and critic networks $\thetaA$ and $\thetaC$ according to \eqref{eq:DDPG_loss_function} and \eqref{eq:DDPG_actor_network_sampled_policy_gradient}.
\item Update the target actor and critic networks $\targthetaA$ and $\targthetaC$ with respect to \eqref{eq:DDPG_target_from_online}. 
\end{enumerate}
We give the pseudocode in \algref{alg:DRL-policy} and the flowchart in \figref{fig:Algorithm-flowchart} for completeness. Note that, as shown in \figref{fig:Algorithm-flowchart}, we are using \algref{alg:threshold-based-heuristic} as the environment with respect to the DDPG agent. 

\begin{algorithm}[]
\caption{DRL-based joint relay selection and beam management strategy}
\label{alg:DRL-policy}
\begin{algorithmic}[1]
\STATE{Input: Length $\horizonlen$ of decision horizon, set $\setofrelayindex$ of relays, minibatch sample size $\batchsize$, replay buffer $\cD$, exploration noise distribution $\cN$, length $\BAlen$ of beam alignment period}
\STATE{Randomly initialize online critic network $Q(s,a|\thetaC)$ and online actor network $\mu(s|\thetaA)$ with $\thetaC$ and $\thetaA$}
\STATE{Initialize target critic network $\targthetaC\leftarrow\thetaC$ and target actor network $\targthetaA\leftarrow\thetaA$}
\FOR{$\horizonindex=1,\ldots,\horizonlen$}
\STATE{Select action $a[\horizonindex]= \{\taurelay[\horizonindex],\taumode[\horizonindex]\}$ according to the current online actor network and exploration noise distribution $\cN$}	
\STATE{Deploy \algref{alg:threshold-based-heuristic} with inputs $\taurelay[\horizonindex]$, $\taumode[\horizonindex]$, $\relayindex[\horizonindex]$, $\modeindex[\horizonindex]$, link vectors $\bb_{0}[\horizonindex],\ldots,\bb_{M}[\horizonindex]$, and $\BAlen[\horizonindex]$.}	
\STATE{Compute reward $r[\horizonindex] = S$ from  \algref{alg:threshold-based-heuristic}}
\STATE{Update $\relayindex[\horizonindex+1]$ and $\modeindex[\horizonindex+1]$ from output of \algref{alg:threshold-based-heuristic} }
\STATE{Get successor state $s[\horizonindex+1]$ from updated link vectors}
\STATE{Store transition $(s[\horizonindex],a[\horizonindex],r[\horizonindex],s[\horizonindex+1])$ in $\cD$}
\STATE{Sample a random minibatch of $\batchsize$ transitions  from $\cD$}
\STATE{Update the online critic network by minimizing the loss \eqref{eq:DDPG_loss_function}}
\STATE{Update the online actor network by policy gradient \eqref{eq:DDPG_actor_network_sampled_policy_gradient}}
\STATE{Update the target networks from the online networks according to \eqref{eq:DDPG_target_from_online} }	
\ENDFOR
\end{algorithmic}
\end{algorithm}

\section{Experimental results} \label{sec:experiments}

In this section, we present the numerical evaluation of the proposed DRL-based algorithm for joint relay selection and beam management problem in a mmWave MIMO vehicular network. We describe the simulation setup and the relevant parameters in \secref{sec:simulation_setup}. We use two scenarios, one to focus on the line-of-sight (LOS) channel and the other to capture non-LOS (NLOS) paths in vehicular networks. We detail the baseline policies and the performance metric in \secref{sec:baseline_algorithms}. We provide the numerical results on the LOS scenario in \secref{sec:numerical_evaluation}. We then give the numerical results on the more realistic scenario with NLOS paths in \secref{sec:experimental_SUMO}.

\subsection{Simulation setup} \label{sec:simulation_setup}

We simulate a mmWave MIMO vehicular network using two scenarios. The first scenario only considers a LOS channel with two relay nodes available to the transmitter. In the second scenario, the channels are calculated using vehicle trajectory data based on Simulator of Urban Mobility (SUMO)~\cite{KraErdBeh:SUMO:12}. The first scenario represents a simplified version of a conceptual deployment for mobile mmWave networks. It is used to analyze the effect of system parameters, such as angular spread $\angularspread$, on the spectral efficiency. The second scenario represents a more realistic deployment of mmWave vehicular networks and is used to analyze specific system parameters of the vehicular network, such as vehicle density per lane, on the spectral efficiency. We assume stationarity in the joint process of channel and blockage in both scenarios, which is commonly used in vehicular channel modeling \cite{GarMolBob:5G_NR_V2X_Tutorial:IEEE21}.

The simulation parameters for both scenarios, unless stated otherwise, are summarized as follows:
\begin{itemize}
\item \textit{Antenna array and codebook:} We assume uniform linear arrays with half-wavelength spacing equipped at both transmitter and receivers. For simplicity of exposition, we focus on a case with uniform linear arrays (ULAs) at the transmitter and receiver, but it can be readily extended to other array geometry and multiple panels. Denoting $\phi$ the steering angle and $\lambda$ the carrier wavelength, the array response vector for a $N$-element ULA is given as 
\begin{IEEEeqnarray}{lCr}
\ba(\phi) = \frac{1}{\sqrt{N}}\left[1, e^{-\jj\pi \cos (\phi)}, \ldots,e^{-\jj(N-1)\pi \cos (\phi)}\right]^{\mathrm{T}}.
\label{eq:array_response_vector}
\end{IEEEeqnarray}
We select a codebook structure that equally partitions the angular domain $[0,\pi]$. The codebook vectors are given as $\bff_{i_{\cF}} = \ba(\pi i_{\cF}/\Nt),$ for $i_{\cF}=0,1,\ldots,\Nt-1$ and similarly for the receiver codebook $\cW$ and the $m$th relay codebook $\cG_{m}$ over $m\in\setofrelayindex$.
\item \textit{Channel model:} We use a time-varying geometric channel composed of $\pathnum[\horizonindex]$ paths as in \cite{VaVikHea:Beam_Tracking_mobile_mmWave:GlobalSIP16}. For the $\pathindex$th path, we denote $\alpha_{\pathindex}[\horizonindex]$ as the complex path gain, $\phinA[\horizonindex]$ as the AOA, $\phinD[\horizonindex]$ as the AOD, $\TXArrayResp(\cdot)$ as the transmit array vector, and $\RXArrayResp(\cdot)$ as the receive array vector. To further express the wideband channel, we apply the delay-$d$ channel model denoting the path delay as $\tau_{\pathindex}$, the bandlimited pulse shaping filter as $p(\cdot)$, the symbol period as $T_{\text{s}}$, and the delay tap length as $\delaytap$ \cite{ParAliHea:Spatial-channel-covariance-estimation-hybrid-architecture-tensor:TCOMM19}. We select $K=256$ subcarriers. We additionally denote the blockage coefficient as $\blockagecoeff$. The channel matrix at subcarrier $\subcarrierindex$ and time slot $\horizonindex$ can be expressed as 
\begin{IEEEeqnarray}{lCr}
\channelmatrix[\subcarrierindex,\horizonindex] = \sum_{\pathindex=1}^{\pathnum[\horizonindex]}\blockagecoeff\alpha_{\pathindex}[\horizonindex]\sum_{d=0}^{\delaytap-1}p(dT_{\text{s}}-\tau_{\pathindex})e^{-\jj \frac{2\pi\subcarrierindex}{\subcarriernum}}\RXArrayResp(\phinA[\horizonindex])\TXArrayResp^{*}(\phinD[\horizonindex]).
\label{eq:channel}
\end{IEEEeqnarray}
We assume that the complex path gain, angle of arrival, and angle of departure evolves according to a first order Gauss-Markov equation, as in \cite[Eq. 7]{VaVikHea:Beam_Tracking_mobile_mmWave:GlobalSIP16}. We denote the angular spread as $\angularspread$, and the complex path gain spread as $\corcoeff$.
\item \textit{Beam management and algorithm initialization:} We apply beam management with $\SSperiodicity=1$ and $\NSS=64$. We assume the transmitter initially uses the direct link and performs initial access. We accordingly initialize the relay index as $\relayindex[1]=0$ and the mode as $\modeindex[1]=0$. We initialize the link vectors as $\bb[1] = \{1,1,0,\ldots,1,1,0\}$. We assume the data transmission takes a single time slot and accordingly set $\DTlen=1$.
\end{itemize}
The parameter values used in both scenarios are organized in \tabref{tab:simulation_setup_parameters}.

\begin{table*}
\caption{Table of the notations, parameters, and values used in the simulation setup in \figref{fig:numerical_performance_comparison}. The following figures have parameter values here unless mentioned otherwise.}\label{tab:simulation_setup_parameters}
\centering
\begin{tabular}{|c|c|c|}
\hline 
Notation & Simulation parameter & Parameter value \\
\hline\hline 
$\nrelay$ & Number of candidate relays & 2 \\\hline
$\Nt$ & Number of transmitter antennas & 16 \\\hline
$\Nr$ & Number of receiver antennas & 16 \\\hline
$\corcoeff$ & Complex path gain spread & 0.005 \\\hline
$\angularspread$ & Angular spread & 0.5 \\\hline
$\blockagelen$ & Number of time slots in a blockage & 100 \\\hline
$T_{\text{s}}$& Symbol time & ${1/1760}$ $\mu$s \\\hline
$\SSperiodicity$ & Number of time slots of a single SS burst & 1 \\\hline
$\NSS$ & Number of SS blocks in single burst & $64$ \\ 	\hline
$p_{\text{u}\rightarrow\text{b}}$ & Transition probability from blocked state to unblocked state & 0.01\\ \hline
$p_{\text{b}\rightarrow\text{u}}$ & Transition probability from unblocked state to blocked state & 0.99\\ \hline
$q_{\text{b}}$ & Steady-state probability for the blocked state & 0.01 \\ \hline
$\subcarriernum$ & Number of subcarriers & 256 \\ \hline
\end{tabular}
\end{table*}

\subsection{Performance metrics and baseline policies} \label{sec:baseline_algorithms}

We use the ensemble average spectral efficiency to track the performance metric. We approximate the ensemble mean by averaging over 1,000 identically distributed channel samples. For the performance of the DRL-based policy, we measure the average of the last 20 iterations out of the $\horizonlen=200$ total iterations to represent the converged reward.  

We compare the proposed DRL-based algorithm to three baseline policies:
\begin{itemize}
\item \textbf{Genie-aided} policy: This algorithm has perfect knowledge of the channel. Subsequently, this policy chooses the data transmission action with the correct relay index and the best beam indices. Therefore, the performance achieved by the genie-aided policy is the expected upper bound of the system.	 
\item \algref{alg:threshold-based-heuristic} with \textbf{optimal threshold}: This algorithm applies \algref{alg:threshold-based-heuristic} with the optimal thresholds $\tau^{\textrm{OPT}}_{\textrm{relay}}$ and $\tau^{\textrm{OPT}}_{\textrm{mode}}$, where $\tau^{\textrm{OPT}}_{\textrm{relay}}$ and $\tau^{\textrm{OPT}}_{\textrm{mode}}$ are found by exhaustively searching over $\taurelay$ and $\taumode$; we return the best result from the tests with varying $\tau_{\textrm{mode}}$ and $\tau_{\textrm{relay}}$ from $0$ up to $\tau_{\max}$ where $\tau_{\max}$ is the 99\% percentile of the achievable spectral efficiency. 
\item \textbf{Direct} policy: This algorithm chooses an action in each iteration following the genie-aided policy and expect the relay index fixed to zero. This policy represents the expected performance using suitable beam tracking and alignment without the aid of available relays.
\end{itemize}

Selected implementation details that may be useful for reproduction are summarized as follows. To implement the proposed learning algorithm based on policy gradients, we use OpenAI Gym \cite{BroChePet:OpenAI-gym:16} as the environment template with Python TensorFlow. We set the action arguments $a_{1}$ and $a_{2}$ as real numbers such that $\taurelay = 10^{a_{1}/10}$ and  $\taumode=10^{a_{1}/10}+10^{a_{2}/10}$; we learn the dB representation of the threshold $\taurelay$ and the dB representation of the difference between thresholds $\taumode-\taurelay$. We found this useful since it allows the $\tanh$ activation function which is known for its stable convergence in training the neural network. An implementation of our method is available on our github page \cite{Kim:Github-Joint-relay-selection-beam-management:22}.

\subsection{Numerical evaluation with LOS channels} \label{sec:numerical_evaluation}

In this section, we provide the experimental results for the scenario that only considers LOS channels between the vehicles. We observe the change in spectral efficiency when varying system parameters. We select the transmit SNR, complex path gain spread $\corcoeff$, angular spread $\angularspread$, codebook size $\codebooksize$, beam management parameters $\NSS$, $\SSperiodicity$, and blockage parameter $q_{\text{b}}$ as the parameters of interest.

\begin{figure}[h!]
\centering
\includegraphics[width=2.5in,draft=false]{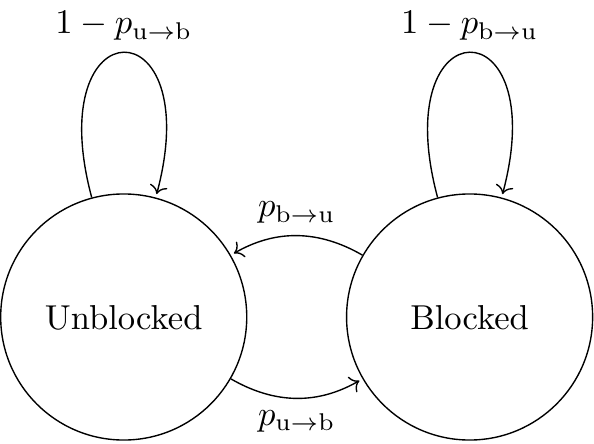}
\caption{LOS blockage evolution model represented as a two-state Markov chain. The steady state probability of blocked state can be computed as $q_{\text{b}} = {p_{\text{u}\rightarrow\text{b}}}/{(p_{\text{b}\rightarrow\text{u}}+p_{\text{u}\rightarrow\text{b}})}$, given the transition probabilities $p_{\text{b}\rightarrow\text{u}}$ from blocked to unblocked state and $p_{\text{u}\rightarrow\text{b}}$ from unblocked to blocked state.}
\label{fig:simplified_LOS_NLOS_switch_model}
\end{figure}

We assume that the time-varying blockage model of the LOS channel scenario can be described by a Markov chain, as in \cite{BobGonXu:LOS-blockage-evolution-V2V-channel:VTC16}. The blockage model, depicted in \figref{fig:simplified_LOS_NLOS_switch_model}, consists of two states indicating the path being blocked or unblocked. We denote the transition probabilities $p_{\text{b}\rightarrow\text{u}}$ from blocked to unblocked state and $p_{\text{u}\rightarrow\text{b}}$ from unblocked to blocked state. The transition probabilities determine the steady-state distribution of the two states. Denoting $q_{\text{u}}$ the steady-state probability of the unblocked state and $q_{\text{b}}$ the steady-state probability of the blocked state, $q_{\text{u}} = \frac{p_{\text{b}\rightarrow\text{u}}}{p_{\text{b}\rightarrow\text{u}}+p_{\text{u}\rightarrow\text{b}}}$ and $q_{\text{b}} = \frac{p_{\text{u}\rightarrow\text{b}}}{p_{\text{b}\rightarrow\text{u}}+p_{\text{u}\rightarrow\text{b}}}$. 
We apply the blockage model along with the evolution of the time-varying propagation channel in \eqref{eq:channel}. We assume that a state transition in the blockage model takes $\blockagelen$ time slots. Typically, $\blockagelen>>1$ since the duration of a blockage is much longer than the symbol period \cite{BobGonXu:LOS-blockage-evolution-V2V-channel:VTC16}. For each path $\pathindex$, $\blockagecoeff=1$ for $\blockagelen$ time slots if the state transits to the unblocked state. If the state transits to the blocked state, $\blockagecoeff=0$ for $\blockagelen$ time slots.

\begin{figure}
\centering
\includegraphics[width=2.5in,draft=false]{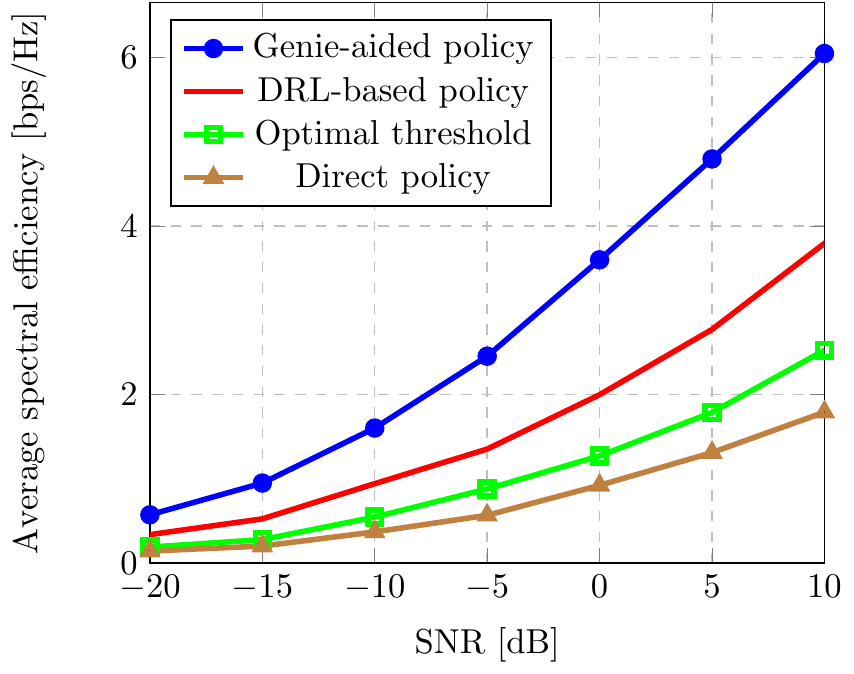}
\caption{Average spectral efficiency vs. transmit SNR for (i) the genie-aided policy, (ii) the DRL-based policy, (iii) the relay selection heuristic with optimal threshold, and (iv) the policy that only use the direct link. Allowing the use of relays improve spectral efficiency overcoming the blockage of LOS path. Relay selection based on DRL further increases spectral efficiency over random selection by balancing exploration and exploitation with $\epsilon$-greedy method.}
\label{fig:numerical_performance_comparison}
\end{figure}

In \figref{fig:numerical_performance_comparison} we illustrate the average spectral efficiency versus SNR, ranging over $-20$ dB to $10$ dB under the parameters specified in \tabref{tab:simulation_setup_parameters}. \figref{fig:numerical_performance_comparison} shows that the proposed learning-based relay selection algorithm achieves spectral efficiency surpassing \algref{alg:threshold-based-heuristic} and the direct policy. This implies that the DRL-based policy is accurately choosing relay indexes to overcome the blockage of the direct LOS path. Furthermore, the DRL-based policy using $\epsilon$-greedy method efficiently balances the tradeoff between spectral efficiency gain from frequent beam alignment and loss from beam management overhead. When compared to \algref{alg:threshold-based-heuristic} using relays, the DRL-based policy achieves non-negligible spectral efficiency increase due to resolving the tradeoff.    

\begin{figure}[t!]
\centering
\subfloat[]{\includegraphics[width=2.5in,draft=false]{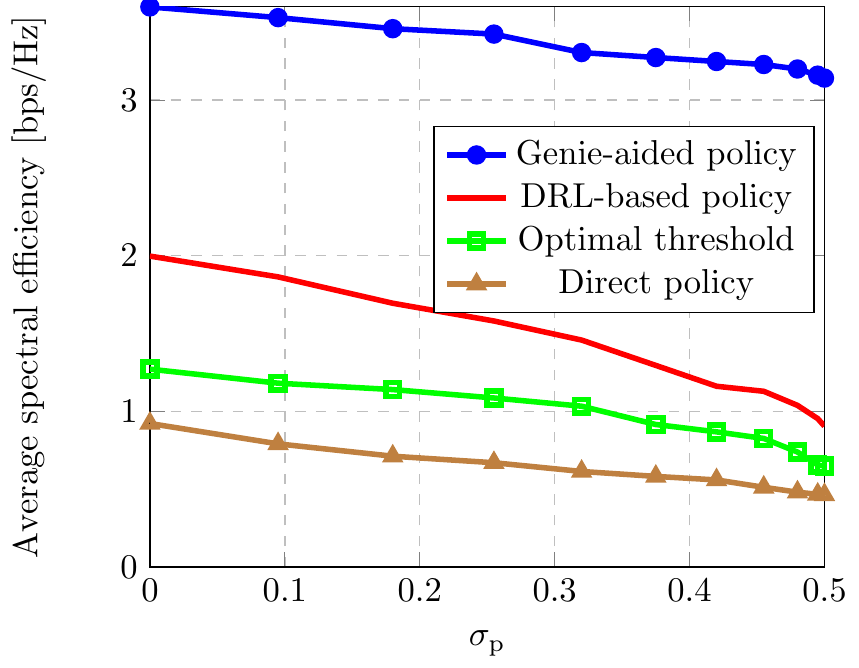}%
\label{fig:correlation_complex_path_gain}
}
\hfil
\subfloat[]{\includegraphics[width=2.5in,draft=false]{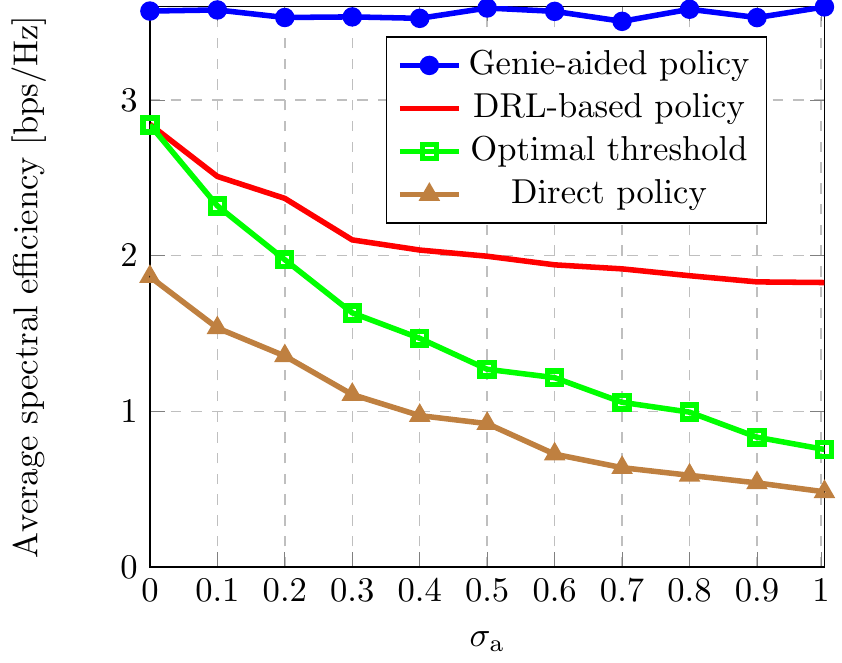}%
\label{fig:angular_variance}
}
\caption{Average spectral efficiency vs. channel parameters (a) complex path gain spread $\corcoeff$ and (b) angular spread $\angularspread$. The DRL-based policy achieves more spectral efficiency compared to the baselines under low complex path gain spread $\corcoeff$. Spectral efficiency achieved by the DRL-based policy degrades slower as the $\angularspread$ increases compared to that of the baseline with prior channel knowledge.
}
\label{fig:effect_of_channel_parameters}
\end{figure}

\figref{fig:effect_of_channel_parameters} illustrates the performance of the policies per channel parameters, complex path gain spread $\corcoeff$ and angular spread $\angularspread$. Low $\corcoeff$ and high $\angularspread$ translates to a fast-varying system with complex traffic; the noise term becomes dominant in the recurrence relations of complex path gain, AOA, and AOD. For fixed SNR at $0$ dB, we vary $\corcoeff$ and $\angularspread$ within $[0,1]$. We fix the angular spread to $0.5$ when varying $\corcoeff$ and we fix the standard deviation of complex path gain noise to $0.005$ when varying $\angularspread$. The DRL-based policy still outperforms \algref{alg:threshold-based-heuristic} and the direct policy for varying $\corcoeff$ and $\angularspread$. We observe interesting behaviors for specific $\corcoeff$ and $\angularspread$ regimes. For instance, the DRL-based policy gain more performance per decreased $\corcoeff$ compared to the baselines. This indicates that the DRL-based policy may be further enhanced with power allocation designs that address variant complex path gain. The performance of the DRL-based policy is resilient against increasing $\angularspread$ compared to that of \algref{alg:threshold-based-heuristic} and direct policy. This implies that the DRL-based policy is particularly beneficial under highly-variant channels.

\begin{figure}
\centering
\includegraphics[width=2.5in,draft=false]{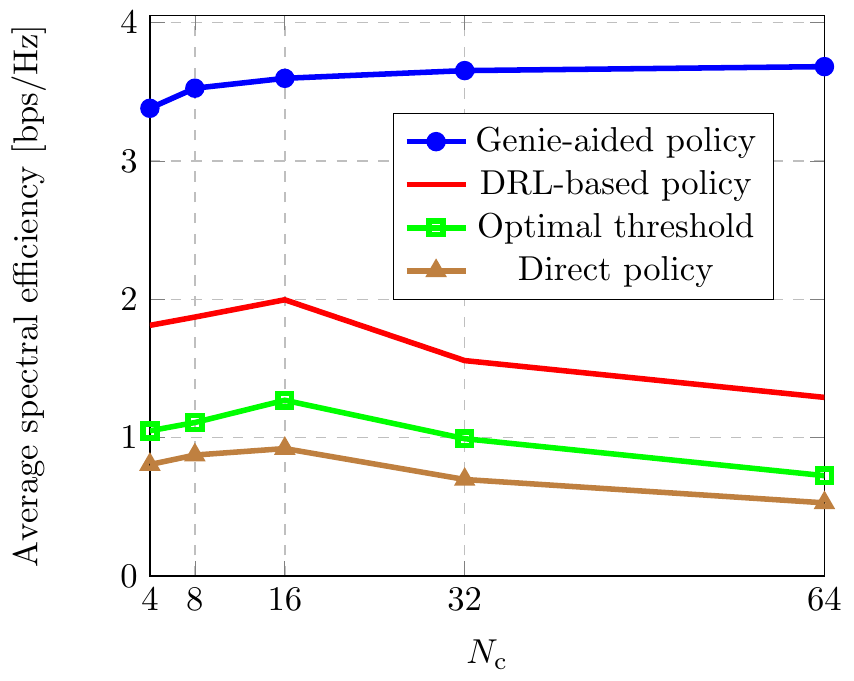}
\caption{Average spectral efficiency vs. transmit SNR for different codebook sizes. The relay codebook sizes and receiver codebook size are set equal to $\codebooksize$. Increasing the codebook size from small $\codebooksize$ results an increase of spectral efficiency due to accurate quantization of the beam angles. For high $\codebooksize$, however, the overhead from beam management dominates the quantization accuracy resulting in a decrease of spectral efficiency.}
\label{fig:effect_of_system_parameters}
\end{figure}

\begin{figure}[!h]
\centering
\subfloat[]{\includegraphics[width=2.5in,draft=false]{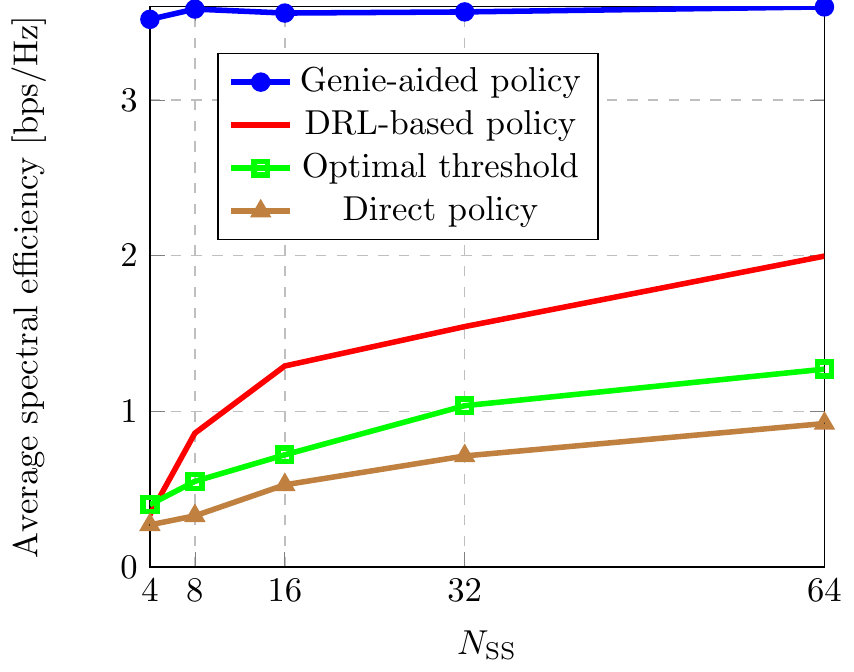}%
\label{fig:number_of_SS_block}
}
\hfil
\subfloat[]{\includegraphics[width=2.5in,draft=false]{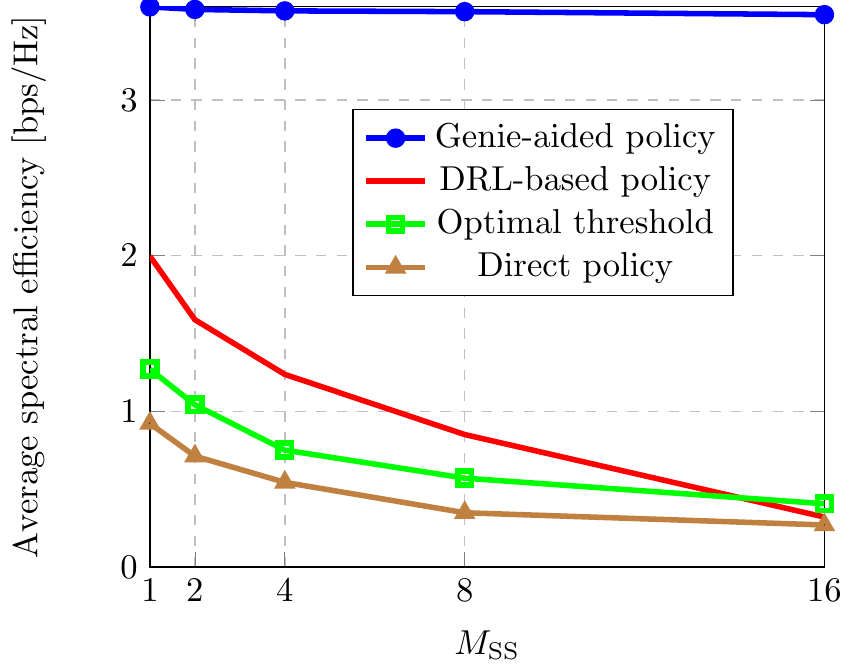}%
\label{fig:SS_burst_periodicity}
}
\caption{Average spectral efficiency vs. different beam management parameters: (a) $\NSS$ and (b) $\SSperiodicity$. Decreasing $\NSS$ and increasing $\SSperiodicity$ results in larger overhead spent in initial access and beam tracking. While the DRL-based policy outperforms the baselines in most $\NSS$ and $\SSperiodicity$ condition, it may underperform under extreme overhead.}
\label{fig:effect_of_beam_management_parameters}
\end{figure}

\figref{fig:effect_of_system_parameters} shows the impact of codebook size on the performance of policies. We vary the codebook size for the transmitter, relay, and receiver from $4$ to $64$ for the 16-element ULA equipment. We observe that increasing the codebook size from $\codebooksize=4$, all strategies gain spectral efficiency. This is expected, since it is known that insufficient quantization of beam angles results in performance degradation for analog beamforming \cite{ElaRajHea:OMP_mmWave_MIMO:TCOM14}. At $\codebooksize=16$, increasing the codebook size results in a decrease of spectral efficiency except for the genie-aided policy. This indicates the spectral efficiency lost in the beam management procedure dominates the spectral efficiency gain from higher beam angle quantization. \figref{fig:effect_of_system_parameters} suggests that there is a codebook size that maximizes the spectral efficiency. While we simulated a codebook equally partitioning the angular domain $[0,\pi]$, it is likely that a similar tradeoff between beam angle quantization and overhead from codebook size exists for other codebooks.

\begin{figure}[t]
\centering
\includegraphics[width=2.5in,draft=false]{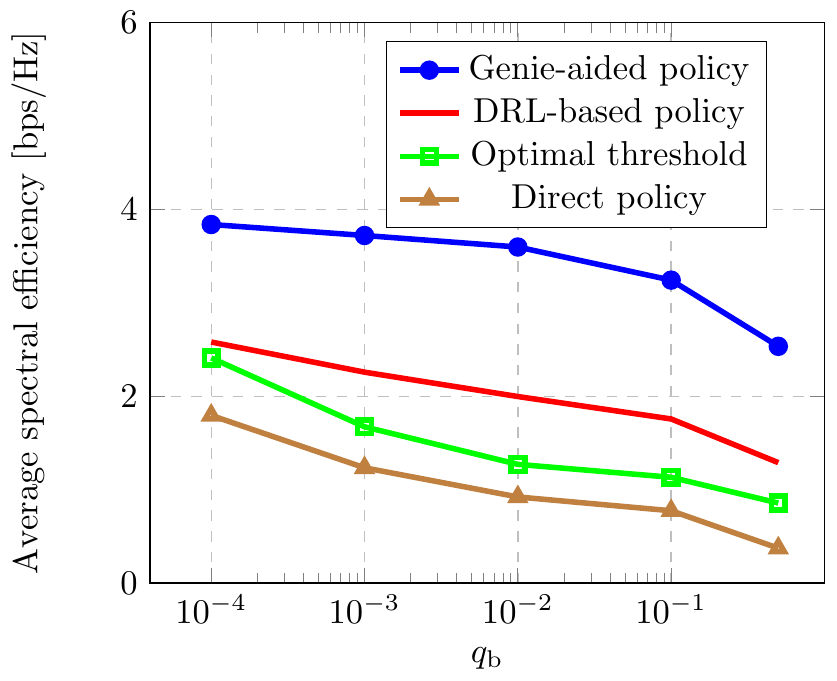}
\caption{Average spectral efficiency vs. different blockage parameter $q_{\text{b}}$. Various blockage parameters $q_{\text{b}}\in\{0.0001,0.001,0.01,0.1,0.5\}$ are plotted to represent the negligible $(q_{\text{b}}<0.01)$, low $(q_{\text{b}}=0.01)$, and high $(q_{\text{b}}=0.5)$ traffic densities. The DRL-based policy shows gradual slope similar to that of genie-aided policy's, which implies that it effectively mitigates blockage similar to the optimal policy.}
\label{fig:effect_of_blockage_parameters}
\end{figure}

In \figref{fig:effect_of_beam_management_parameters} we demonstrate the effect of the parameters related to SS bursts and blocks. We vary the number $\NSS$ of SS blocks per burst in $\{8,16,32,64\}$ and periodicity $\SSperiodicity$ of SS bursts in $\{1,2,4,8,16\}$, as in  \cite{GioPolZor:Beam_Management_3GPP_NR_mmWave_Tutorial:COMM18}. \figref{fig:effect_of_beam_management_parameters} shows that the DRL-based policy outperforms baselines in most cases but it may underperform when $\NSS$ is low or $\SSperiodicity$ is high. For example, the DRL-based policy severely lose performance both at $\NSS=4$ and $\SSperiodicity=16$. Such low performance of the DRL-based algorithm happens because the increased time slots required for exploration causes the learning algorithm to fail to converge. This implies that the DRL-based policy is sensitive to beam management parameters, but it works well under practical scenarios.

\figref{fig:effect_of_blockage_parameters} illustrates the effect of the blockage parameter. We vary the steady-probability $q_{\text{b}}$ of blocked state in $\{0.0001,0.001,0.01,0.1,0.5\}$. For a given $q_{\text{b}}$, we use a Markov chain in \figref{fig:simplified_LOS_NLOS_switch_model} with transition probabilities set to $p_{\text{u}\rightarrow\text{b}}=q_{\text{b}}$ and $p_{\text{b}\rightarrow\text{u}}=1-p_{\text{u}\rightarrow\text{b}}$. We simulate the scenario with a high vehicular density by setting $q_{\text{b}}=0.5$, low density by setting $q_{\text{b}}=0.01$, and negligible density by setting $q_{\text{b}}<0.01$. \figref{fig:effect_of_blockage_parameters} depicts that DRL-based policy behaves similarly to the genie-aided policy over the change of $q_{\text{b}}$ compared to baselines. Both baselines severely lose spectral efficiency compared to the DRL-based policy within the negligible density regime, whereas the genie-aided policy suggests high spectral efficiency can be maintained within the negligible density regime. This implies that the DRL-based policy is able to effectively mitigate blockage by jointly selecting the relay and the mode. 

\subsection{Numerical evaluation on SUMO-generated channel}\label{sec:experimental_SUMO}

In this section, we provide the experimental results for the scenario that represents a more realistic deployment of a mmWave MIMO vehicular network. We follow the approach in \cite{KlaBatGon:5GMdata:18} to generate the channels based on the time-varying wideband channel \eqref{eq:channel} and the vehicle trajectories from SUMO. We apply a simple ray tracing method to obtain the number of paths $\pathnum[\horizonindex]$ and blockage coefficient $\blockagecoeff$ assuming all vehicles have length of $4.645$ m, vehicles can block LOS, and the vehicle surfaces act as lossless reflectors to create reflected paths. We calculate the AOA/AOD and path gain assuming the ray propagation starts at the end of vehicles facing each other, the angle of the reflected ray by the vehicle surface is equal to the angle of incident ray, and the path loss exponent is 2. We report the change in spectral efficiency when varying system parameters. We select the transmit SNR, vehicle density, and average vehicle speed as the parameters of interest. 

\begin{figure}[t]
\centering
\includegraphics[width=2.5in,draft=false]{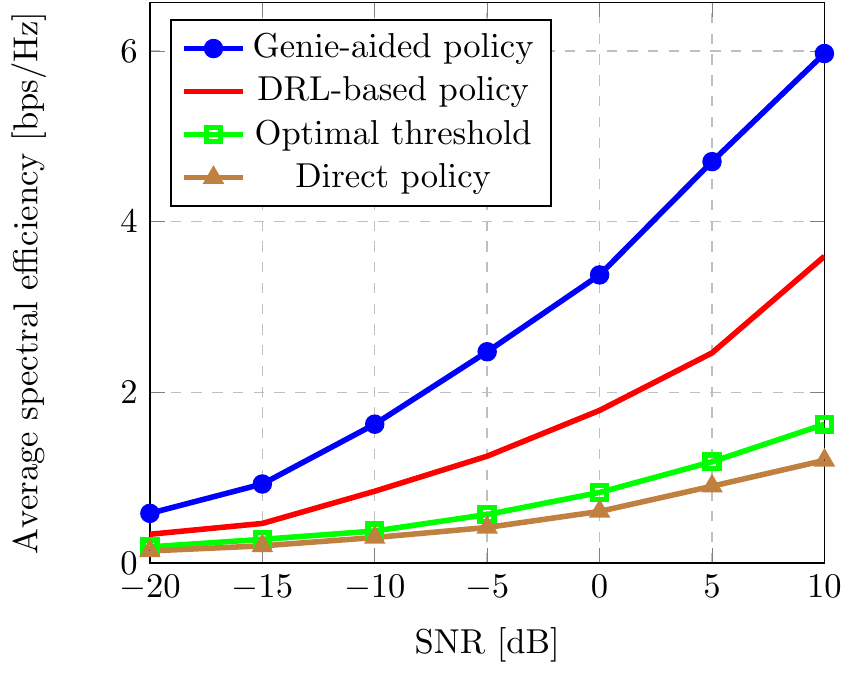}
\caption{Average spectral efficiency vs. transmit SNR for (i) the genie-aided policy, (ii) the DRL-based policy, (iii) the relay selection heuristic with optimal threshold, and (iv) the policy that only use the direct link. Similar to that observed in \figref{fig:numerical_performance_comparison}, the proposed DRL-based policy improves spectral efficiency over baseline methods.}
\label{fig:SNR_SE_SUMO}
\end{figure}

In \figref{fig:SNR_SE_SUMO} we show the average spectral efficiency versus SNR, ranging over $-20$ dB to $10$ dB under the parameters specified in \tabref{tab:simulation_setup_parameters}. We set the traffic density as 10 vehicles per km and the average vehicle speed as 80 km/h. \figref{fig:SNR_SE_SUMO} confirms that the proposed DRL-based relay selection policy outperforms baselines in a realistic scenario.
We observe the spectral efficiency obtained using \algref{alg:threshold-based-heuristic} is closer to the spectral efficiency using the direct policy decrease compared to that in \figref{fig:numerical_performance_comparison}. This highlights the model-free aspect of the proposed DRL algorithm, which may further outperform policies based on fixed data in realistic scenarios due to the increased model complexity.

\begin{figure}[t]
\centering
\includegraphics[width=2.5in,draft=false]{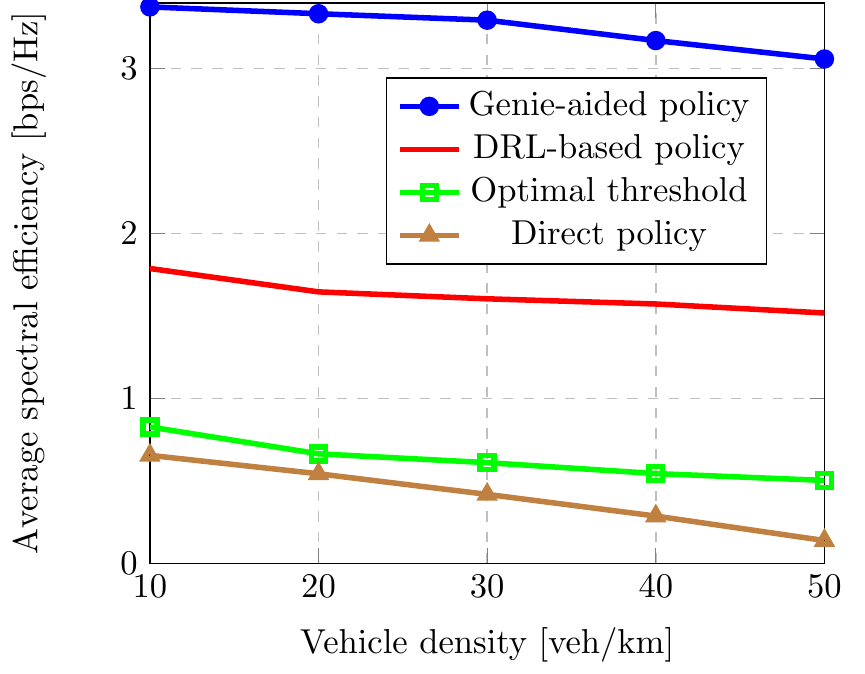}
\caption{Average spectral efficiency vs. different vehicle densities. Overall policies suffer spectral efficiency loss due to the increased chance of blockage from higher vehicle density. Still, the proposed DRL-based policy outperforms baselines, especially under dense vehicle networks, by efficiently using the indirect links to avoid the frequent blockage of the LOS paths.}
\label{fig:effect_of_vehicle_density}
\end{figure}

\figref{fig:effect_of_vehicle_density} shows the effect of vehicle density. We vary the number of vehicles per kilometer from 10 to 50 in the SUMO simulation. We observe a loss spectral efficiency achieved by the proposed DRL-based policy as the vehicle density increases. Still, the performance loss of the DRL-based policy due to the increase in the vehicle density is minor compared to that of direct policy, which plummets in the congested case. Since the direct policy only uses the direct link, this indicates that cooperative relays become more beneficial
as the vehicular networks gets denser.

\begin{figure}[t]
\centering
\includegraphics[width=2.5in,draft=false]{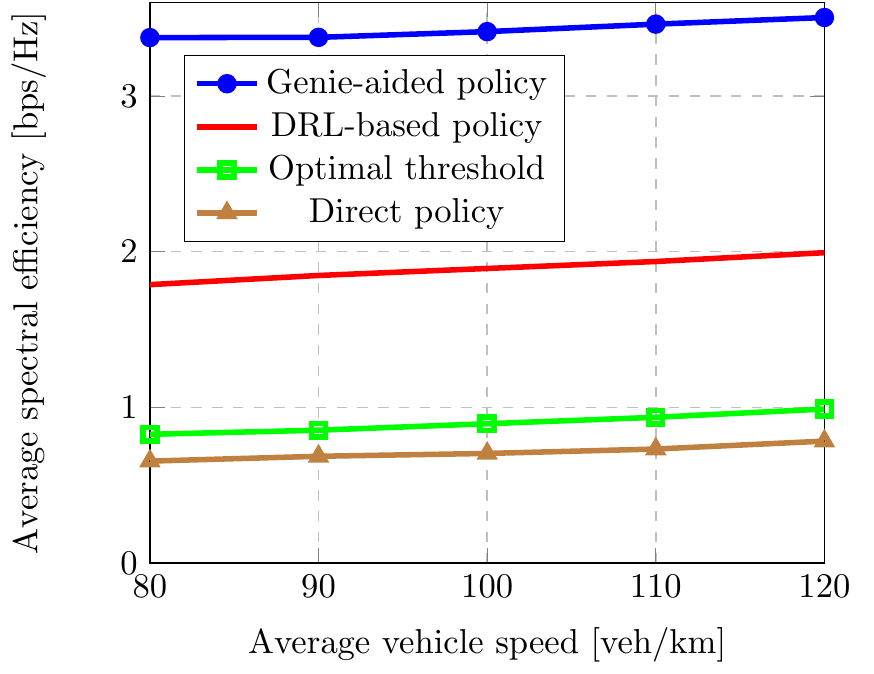}
\caption{Average spectral efficiency vs. average vehicle speeds. Increased mobility, which may decrease the blockage duration, shows an overall increase in spectral efficiency for all of the considered policies. The proposed DRL-based policy outperforms baselines, especially under highly mobile networks.}
\label{fig:effect_of_vehicle_speed}
\end{figure}

\figref{fig:effect_of_vehicle_speed} depicts the impact of average vehicle speed. We select the range of vehicle speed from 80 km/h to 120 km/h, following the common highway speed limit in the United States. The spectral efficiency of all the policies gradually improves as the average vehicle speed increases. The performance enhancement may be due to the decreased blockage duration from the increased vehicle speed, despite negative performance factors such as increased beam alignment frequencies \cite{TunOzkPan:Latency-mmWave-connected-vehicular-network:21VT}. The proposed DRL algorithm shows the steepest increase of spectral efficiency compared to the baselines. \figref{fig:effect_of_vehicle_speed} indicates that proposed relay selection algorithm is suitable for mobile vehicular networks, especially those with high mobility.

\section{Conclusions and future work} \label{sec:conclusion}

Future vehicular networks will benefit from relay selection algorithms addressing the frequent blockages induced by dense deployment of mobile nodes. Regarding the higher frequency bands used at 5G at beyond, sources of overhead should be incorporated in the analysis of relay selection algorithms. We derived an MDP and devised a DRL-based algorithm for the spectral efficiency optimization problem accounting both relay selection and beam management. We observed that the spectral efficiency achieved by the proposed method is greater than that of a fixed threshold policy over different transmit SNRs. The simulation results show that the DRL-based algorithm can adapt to fast-varying channels using beam measurements, which are compared with thresholds, to determine actions. This indicates the proposed DRL algorithm can be implemented to vehicular networks to maximize spectral efficiency by exploiting the time-varying adaptive thresholds. For future work, we plan to extend our work to incorporate fast beam alignment algorithms and realistic beam measurement feedback procedures.

\bibliography{main_v24}	
\bibliographystyle{IEEEtran}

\end{document}